\documentclass[pre,preprint,amsmath,amssymb,showpacs]{revtex4-1}

\usepackage{dcolumn}
\usepackage{bm}
\usepackage[dvips]{color}
\usepackage{subfigure}
\usepackage{graphicx} 
\usepackage{algorithmicx}
\usepackage[ruled]{algorithm}
\usepackage{algpseudocode}

\newcommand{\Amat}{\textbf{A} }

\newcommand{\Uvec}{\textbf{\textit{U}} }
\newcommand{\dif}{{\rm d} }

\begin{document}

\title{Collective Dynamics of Interacting Particles in Unsteady Flows} 

\author{Maryam Abedi}
\author{Mir Abbas Jalali}
\email{mjalali@sharif.edu}

\affiliation{
Computational Mechanics Laboratory, Department of Mechanical 
Engineering, Sharif University of Technology, Azadi Avenue, P.O. Box: 11155-9567, Tehran, Iran}

\begin{abstract}
We use the Fokker-Planck equation and its moment equations to study the collective behavior
of interacting particles in unsteady one-dimensional flows. Particles interact according to a 
long-range attractive and a short-range repulsive potential field known as Morse potential.  
We assume Stokesian drag force between particles and their carrier fluid, and find analytic
single-peaked traveling solutions for the spatial density of particles in the catastrophic phase. 
In steady flow conditions the streaming velocity of particles is identical to their carrier fluid, 
but we show that particle streaming is asynchronous with an unsteady carrier fluid. Using 
linear perturbation analysis, the stability of traveling solutions is investigated in unsteady 
conditions. It is shown that the resulting dispersion relation is an integral equation of the 
Fredholm type, and yields two general families of stable modes: singular modes whose 
eigenvalues form a continuous spectrum, and a finite number of discrete global modes. 
Depending on the value of drag coefficient, stable modes can be over-damped, critically 
damped, or decaying oscillatory waves. The results of linear perturbation analysis are 
confirmed through the numerical solution of the fully nonlinear Fokker-Planck equation.    
\end{abstract}


\maketitle

\section{Introduction}

Collective motion of interacting particles is observed in various natural systems, 
from large-scale schools of fishes \cite{Cam12} and the flock of birds \cite{Dar09} 
to small-scale aggregation of red blood cells (RBC) \cite{Bor03} and bacterial 
colonies \cite{Chen12}. To elucidate the mechanisms leading to formation of this 
self organized patterns, several particle based \cite{vic95,Ors06}, continuum kinetic 
\cite{Ha08} and hydrodynamic models \cite{for11} have been developed. In these 
systems, the population organization is affected by the influence zone \cite{Agu11},
leadership \cite{She08}, geometrical constraints \cite{Agu11} and also by environmental 
factors like the drag force \cite{Ors06} and random noise in particle velocities \cite{Bag09}. 

The organized movement of particles becomes more complex if they move in a 
transient carrier fluid. For instance, the collaborative flock of birds in windy conditions, 
and the swim of fishes in rivers or along oceanic currents is poorly understood, 
and we know a little about the stability of their group behavior in time-varying 
flow fields. The collective dynamics of RBCs in blood vessels is also expected to 
correlate with the chaotic nature of the heart beat \cite{BD88} and its influence on 
the blood stream. Moreover, micro-organisms produce different collective behavior 
in the presence of turbulence \cite{Pet02}, vortices \cite{Mar06} and shear flow \cite{Dur09}. 
Hydrodynamic models of self-propelled, non-interacting particles in an incompressible 
fluid show that a concentrated population of swimming bacteria in an incompressible 
fluid exhibits spatiotemporal patterns \cite{Che12,ped92}. Other works in the literature have only addressed 
the swarm of interacting particles in the absence of inertial effects and when the carrier 
fluid has a constant velocity \cite{Lev09,bre12,Hac12}. It is therefore important to 
understand how particle--particle interactions and the drag force of an unsteady 
carrier fluid can collaborate to develop patterns in the spatial distribution of 
particles, and whether such patterns can remain stable. 

The flock/swarm pattern depends on the interaction between particles. A widely 
used interaction model is the Morse potential, which can produce various collective 
behaviors, including localized flocks and vortex solutions, in one- and two-dimensional 
systems \cite{Ors06,ber10,Lev01}. The parameters of the Morse potential control the 
phase, catastrophic or $H$-stable, and morphology of self-organized systems. 
A system consisting of $N$ interacting particles and with the total potential 
energy $U$, is in $H$-stable phase if the quantity $U/N$ has a lower negative 
bound, and the system does not collapse in the limit of $N\rightarrow \infty$. 
In such conditions the swarm size typically increases with the number of particles. 
Non-$H$-stable systems with particles collapsing into a dense body are called 
catastrophic \cite{Ru69}. 
Here, we adopt the Morse potential and study the dynamics of colonies of particles in
one-dimensional flows. The Fokker-Planck equation is used to trace the dynamics 
of particles as they sense the unsteady motion of carrier fluid through Stokesian drag. 
The response of flocking particles to disturbances in the density and streaming velocity, 
and the effect of drag coefficient on stability characteristics, are investigated for the 
first time.

The paper is organized as follows. In section \ref{sec:Kinetic_Eq}, we introduce 
our model and derive the kinetic and hydrodynamic equations using the motion 
equations of individual particles. In section \ref{sec:Numerical-methods}, we explain 
numerical algorithms that have been used to solve the Fokker-Planck equation. 
The collective dynamics of particles in steady flows is studied in section \ref{sec:collective-dynamics} 
and exact analytical solutions are found for catastrophic swarms of particles. 
In section \ref{sec:unsteady-flows}, we present analytical solutions for the streaming 
of particles in unsteady flow conditions, and perturb the continuity and momentum 
equations to investigate the linear stability of these time-varying collective motions. 
The results of the linear perturbation analysis are verified using numerical solutions 
of the Fokker-Planck equation. Our concluding remarks are presented 
in section \ref{sec:conclusions}.

\section{Kinetic and Hydrodynamic equations}
\label{sec:Kinetic_Eq}
We assume that particles move in a Newtonian fluid and the drag force exerted on 
the $i$th particle, with the position $x_i$ and velocity $v_i$, is computed from 
$-\beta(v_i-u_{\rm f})$. Here $u_{\rm f}(x,t)$ is the streaming velocity of the carrier 
fluid, and $\beta$ is the Stokesian drag coefficient. The equations of motion for the 
$i$th particle read 
\begin{eqnarray}
\dot{x}_{i} = {v}_{i},~~
\dot{v}_{i} = -\frac{1}{N}\sum_{j=1}^N \partial_{x_{i}} \Phi(r_{ij} ) - \beta(v_{i}-u_{\rm f}),
~~i=1,2,\ldots,N,
\label{eq:dot-xv}
\end{eqnarray} 
with $r_{ij}=\vert x_i - x_j \vert$ and $\partial_{x_i} = \partial/\partial x_i$. 
$\Phi$ is the two-body Morse potential defined as 
\begin{eqnarray}
\Phi(r_{ij}) = c_{1} e^{-r_{ij}/ d_1} - c_{2} e^{r_{ij}/d_2},
\end{eqnarray}
which has been widely used to describe red blood cell aggregation \cite{liu06} and 
the swarm of animals \cite{top08}. The positive parameters $d_2$ and $d_1$ are, respectively, 
the attraction and repulsion length scales, and $c_1>0$ and $c_2>0$ are force magnitudes. 
Long-range attractive, and short-range repulsive forces are obtained by setting 
$d_1/d_2 < 1$ and $c_1/c_2 > 1$. 

Let $f^{(N)}\left (\{ x_{i}\},\{ v_{i}\},t \right )$ ($i=1,2,\ldots,N$) be the $N$-particle probability 
distribution function (DF) at time $t$. The probability of finding the $i$th particle at the 
position $x_{i}$ with the velocity $v_{i}$ within an infinitesimal phase space volume 
$\dif {\cal V}=\Pi_{i=1}^{N}\dif x_{i}\dif v_{i}$ is $f^{(N)}\dif {\cal V}$. Since the mass of the 
entire $N$-particle system is conserved, the temporal evolution of $f^{(N)}$ is governed 
by Liouville's equation \cite{Car09}:
\begin{eqnarray}
\label{eq:kinetic1}
\frac{\partial f^{(N)}}{\partial t}+\sum_{i=1}^{N}\left[ \partial_{x_i} \left ( \dot{x_{i}}f^{(N)} \right )
+ \partial_{v_i} \left (\dot{v_{i}}f^{(N)} \right ) \right]=0,
\end{eqnarray}
The one-particle DF is obtained by integrating $f^{(N)}$ as
\begin{eqnarray}
f^{(1)}(x_1,v_1,t)=\int_{{\cal V}_1} f^{(N)} (x_{1},x_{2},\ldots,x_{N},v_{1},v_{2},\ldots,v_{N},t) 
\, \dif {\cal V}_1,
\label{eq:define-f1}
\end{eqnarray}
where ${\cal V}_1$ is an $(N-2)$ dimensional subspace of ${\cal V}$. The evolutionary 
equation of $f^{(1)}$ can therefore be obtained through integrating equation (\ref{eq:kinetic1}) as
\begin{eqnarray}
\label{eq:kinetic2}
\frac{\partial f^{(1)}}{\partial t} +\int_{{\cal V}_1} \left[ \partial_{x_1} \left (\dot{x_{1}}f^{(N)} \right )+
\partial_{v_1} \left (\dot{v_{1}}f^{(N)} \right ) \right] \, \dif {\cal V}_1=0.
\label{eq:evolution-f1}
\end{eqnarray}
Substituting from (\ref{eq:dot-xv}) into (\ref{eq:evolution-f1}) gives
\begin{eqnarray}
\label{eq:kinet}
\frac{\partial f^{(1)}}{\partial t} + v_1 \frac{\partial f^{(1)}}{\partial x_{1}}
+\frac{\partial}{\partial v_{1}} \left [ -\beta(v_1-u_{\rm f})f^{(1)} \right ]+{\cal A}=0,
\end{eqnarray}
with
\begin{eqnarray}
{\cal A}=-\frac{1}{N}\sum_{j=2}^N \int_{{\cal V}_1} \partial_{v_1} \left [ f^{(N)}\partial_{x_1}\Phi(r_{1j}) \right ] \, \dif {\cal V}_1.
\end{eqnarray}
All integrals within the summation are identical: they are evaluated over $(N-2)$ dimensional subspaces of 
${\cal V}$, and there are $N-1$ of such integrals/subspaces. One thus finds 
\begin{eqnarray}
{\cal A} = -\frac{N-1}{N} \int_{{\cal V}_1} \partial_{v_1} \left [ f^{(N)}\partial_{x_1}\Phi(r_{12}) \right ] \, \dif{\cal V}_1 
= -\frac{N-1}{N} \int \partial_{v_1} \left [f^{(2)}\partial_{x_1}\Phi(r_{12}) \right ] \, \dif x_2 \dif v_2,
\label{eq:A-integrals}
\end{eqnarray}
where the two-particle DF $f^{(2)}(x_1,v_1,x_2,v_2,t)$ is obtained by integrating $f^{(N)}$ over an $(N-4)$ dimensional 
subspace of ${\cal V}$. In a system with long-range interactions, when the relaxation time becomes sufficiently large 
in the thermodynamic limit $N\rightarrow\infty$, two-particle correlations are neglected and one can work with 
the separable form \cite{Car09}
\begin{eqnarray}
f^{(2)}(x_1,x_2,v_1,v_2,t)=f^{(1)}(x_1,v_1,t) f^{(1)}(x_2,v_2,t).
\label{eq:define-f2}
\end{eqnarray}
Substituting from equation (\ref{eq:define-f2}) into (\ref{eq:A-integrals}) leads to
\begin{eqnarray}
{\cal A}=-\frac{N-1}{N} \int \frac{\partial}{\partial v_1} \left [ f^{(1)}(x_1,v_1,t)f^{(1)}(x_2,v_2,t)\partial_{x_1}\Phi(r_{12}) \right ] \, \dif x_2 \dif v_2.
\label{eq:A-final-form}
\end{eqnarray}
In the limit of $N\rightarrow\infty$, the fraction $(N-1)/N$ tends to unity. Defining $f(x,v,t)=f^{(1)}(x,v,t)$ and using 
(\ref{eq:A-final-form}), the evolutionary equation (\ref{eq:kinet}) of the one-particle DF $f$ transforms to the 
Fokker-Planck equation
\begin{eqnarray}
\frac{\partial{f}}{\partial{t}}+v\frac{\partial{f}}{\partial{x}}-\left ( \partial_x \Phi *\rho \right ) 
\frac{\partial{f}}{\partial{v}} = -\frac{\partial}{\partial v} \left ( D[\Delta v] f \right ), 
\label{eq:kinetic}
\end{eqnarray}
where $D[\Delta v]=-\beta(v-u_{\rm f})$ is the diffusion coefficient corresponding to steady 
drift in the phase space. In deriving equation (\ref{eq:kinetic}), we have ignored the random 
motions of particles due to thermal fluctuations. This restricts the applications of our results
to flows with P\'eclet number $\gg 1$. 

The macroscopic density $\rho(x,t)$ and the streaming velocity $u(x,t)$ of particles are defined as
\begin{eqnarray}
\rho(x,t) = \int_{-\infty}^{+\infty} f(x,v,t) \, \dif v,~~
u(x,t) = \frac{1}{\rho(x,t)} \int_{-\infty}^{+\infty} v \, f(x,v,t) \, \dif v,
\end{eqnarray}
and the force field due to particle--particle interactions is computed from the convolution 
integral
\begin{eqnarray}
\partial_x \Phi *\rho=\int \partial_x \Phi( \vert x-y \vert ) \, \rho(y) \, \dif y.
 \end{eqnarray}
Our simulations of (\ref{eq:kinetic}) show that the distribution function acquires a spiky 
nature in the velocity space. Therefore, it is legitimate to work with a monokinetic velocity 
distribution of the form $f(x,v,t)=\rho(x,t)\delta(v-u(x,t))$. Numerical calculations that 
support this assumption will be presented in section \ref{sec:collective-dynamics}. 
Taking the zeroth- and first-order moments of (\ref{eq:kinetic}) gives the continuity and momentum 
equations of the particle phase as
\begin{eqnarray}
\label{eq:hydrodynamic}
\frac{\partial \rho}{\partial t} &+& \frac{\partial}{\partial x}(\rho \, u)=0, \label{eq:hydro-mass} \\
\frac{\partial u}{\partial t} &+& u \, \frac{\partial u}{\partial x} = - \beta(u-u_{\rm f})- \partial_x \Phi *\rho.
 \label{eq:hydro-momentum}
\end{eqnarray}
We assume that the carrier fluid is incompressible. Its dynamics will therefore be governed by 
the following continuity and momentum equations:
\begin{eqnarray}
\frac{\partial u_{\rm f}}{\partial x} &=& 0,  \label{eq:continuity-carrier} \\
\rho_{\rm f} \left ( \frac{\partial u_{\rm f}}{\partial t}+u_{\rm f}\frac{\partial u_{\rm f}}{\partial x} \right )
&=& -\frac{\partial p_{\rm f}}{\partial x}+\mu_{\rm f}\frac{\partial ^2 u_{\rm f}}{\partial x ^2}+\beta \rho (u-u_{\rm f}),
\label{eq:momentum-carrier}
\end{eqnarray} 
with $\rho_{\rm f}$, $\mu_{\rm f}$ and $p_{\rm f}$ being the density, viscosity and pressure 
of the carrier fluid, respectively. Equation (\ref{eq:continuity-carrier}) implies that $u_{\rm f}$ 
is only a function of $t$, and it is prescribed at the inlet of the 
one-dimensional flow. Therefore, any change in the spatial density and streaming velocity 
of the particle phase will locally affect $p_{\rm f}$, which can be computed as the only unknown 
of the momentum equation (\ref{eq:momentum-carrier}). We remark that this conclusion does 
not apply to two- or three-dimensional flows because deformations of streamlines in higher 
dimensions prevents us from solving the continuity equation independently.

\section{Numerical methods}
\label{sec:Numerical-methods}

Equation (\ref{eq:kinetic}) can be solved using spectral methods \cite{Gib06}, particle in cell (PIC) 
methods \cite{jac06}, Fourier transform of equations in the velocity subspace \cite{Eli11}, and 
time-splitting method in combination with finite element or finite volume methods \cite{Car07,Qiu10}. 
In this paper we have used the last method because of its stability in long-time simulations. 
Our experiments show that spectral methods are faster, but the results diverge over short 
time scales. We did not choose PIC methods either because they generate numerical noise,
which must be avoided in stability analysis. 

We perform time-splitting to reduce the Fokker-Planck equation to two one-dimensional advection 
problems:
\begin{eqnarray}
&{}& \frac{\partial{f}}{\partial{t}}+v\frac{\partial{f}}{\partial{x}}=0, \label{eq:split1} \\
&{}& \frac{\partial{f}}{\partial{t}}-\left ( \partial_x \Phi *\rho \right ) 
\frac{\partial{f}}{\partial{v}} +\frac{\partial}{\partial v} \left [ \beta (v-u_{\rm f}) f \right ]=0,
\label{eq:split2}
\end{eqnarray} 
which can be solved using flux balance method (FBM) in combination with point-wise weighted 
essentially non-oscillatory (PWENO) interpolation method. FBM guarantees the conservation of 
mass, and PWENO interpolation avoids spurious oscillations in computing the higher-order 
derivatives of the DF $f$. A detailed description of FBM and PWENO algorithms can be found 
in \cite{Car07}. The discretized forms of equations (\ref{eq:split1}) and (\ref{eq:split2}) are 
\begin{eqnarray}
\frac{\partial f_{1,j}(x,t)}{\partial t}+\frac{\partial }{\partial x} \left [ k_{1,j}(x,t)f_{1,j}(x,t) \right ] &=& 0,~ f_{1,j}(x,t)=f(x,v_j,t),~j=1,\ldots,N_2,
\label{eq:discrete-splitb1} \\
\frac{\partial f_{2,i}(v,t)}{\partial t}+\frac{\partial }{\partial v} \left [ k_{2,i}(v,t)f_{2,i}(v,t) \right ] &=& 0,~ f_{2,i}(v,t)=f(x_i,v,t),~i=1,\ldots,N_1,
\label{eq:discrete-splitb2}
\end{eqnarray}
where we have used uniform grids in both the $x$- and $v$-subspaces and
\begin{eqnarray}
k_{1,j}(x,t) = v_j, ~
k_{2,i}(v,t) = - \left [ \partial_x \Phi *\rho \right ]_{(x_i,t)}+\beta (v-u_{\rm f}).
\end{eqnarray}
Having $f(x_i,v_j,t_n)$ at all grid points at the $n$th time step $t_n$, $f$ is updated at 
$t_{n+1}=t_n+\Delta t_n$ after three successive steps that utilize FBM and PWENO algorithms:
(i) For $1\le j \le N_2$, equation (\ref{eq:discrete-splitb1}) is solved in the $(x,t)$-space over the 
half time step $\Delta t_n/2$. (ii) The system of $N_1$ equations (\ref{eq:discrete-splitb2}) is solved 
in the $(v,t)$-space over a full time step $\Delta t_n$. (iii) Step (i) is repeated for another half time 
step $\Delta t_n/2$. This procedure assures a second-order accuracy in the time domain \cite{che76}. 

In our simulations we assume the periodic boundary condition $f(0,v,t)=f(X,v,t)$ in the 
$x$-direction with $X$ defining the domain of $x\in[0,X]$. The parameter $X$ is estimated based 
on the flock length $L$ that we introduce in section \ref{sec:analytical} and find using steady state
analytical solutions. Boundary conditions in the velocity space are based on excluding escape solutions: 
the number density of particles should vanish for $v \rightarrow \pm \infty$. In numerical calculations 
we cannot reach this very limit, and therefore, use a sufficiently large cutoff speed $V>0$ so that 
$f(x,v,t)=0$ for $v\geq+V$ and $v\leq-V$. 

To test the convergence of numerical simulations, we 
start with an initial rectangular $N_1 \times N_2$ mesh ${\cal M}_1$ in the $(x,v)$-space, and compute 
$f$ and store it at the grid (nodal) points. We then use a second finer mesh ${\cal M}_2$ of the size 
$2N_1 \times 2N_2$, compute $f$ at the grid points of ${\cal M}_1$ by interpolating the nodal values 
of $f$ in ${\cal M}_2$. Comparing the first and second sets of nodal DFs, yields the distribution of 
the computational error in the $(x,v)$-space. According to our numerical experiments, by taking 
$N_1 \times N_2 = 500\times 400$ the maximum relative error is within $2$ percent, which shows 
a reasonable convergence of the FBM and PWENO methods. \\

\section{Collective dynamics in steady flows}
\label{sec:collective-dynamics}

\subsection{Simulation of the Fokker-Planck equation}

Solving equation (\ref{eq:kinetic}) is simpler when $u_{\rm f}$ is constant. In such conditions we identify 
catastrophic and $H$-stable phases \cite{Ors06,Lev09} for the collective dynamics of particles. Consider 
the following transformations: $\tilde{v}=v-u_{\rm f}$ and $\xi = x-u_{\rm f} t$, and define 
$f(x(\xi,t),v(\tilde v),t)=\tilde{f}(\xi,\tilde{v},t)$. The kinetic equation (\ref{eq:kinetic}) thus becomes
\begin{eqnarray}
\frac{\partial{\tilde{f}}}{\partial{t}}+\tilde{v}\frac{\partial{\tilde{f}}}{\partial{\xi}}-\left ( \partial_\xi \Phi *\rho \right ) 
\frac{\partial{\tilde{f}}}{\partial{\tilde{v}}} = -\frac{\partial}{\partial \tilde{v}} \left ( \beta \tilde{v} \tilde f \right ), 
\end{eqnarray}
which means the solutions of the Fokker-Planck equation in coordinate frames moving with 
the constant speed $u_{\rm f}$ can always be mapped to solutions in stationary frames. 
Therefore, if we find a solution $f(x,v,t)$ for $u_{\rm f}=0$, the function 
$\tilde f(\tilde x,\tilde v,t)=f(\tilde x-u_{\rm f}t,\tilde v-u_{\rm f},t)$ will be a solution of 
(\ref{eq:kinetic}) in the $(\tilde x,\tilde v,t)$ space for any $u_{\rm f}\not = 0$. 

\begin{figure*}
\centerline{\hbox{\includegraphics[width=0.4\textwidth]{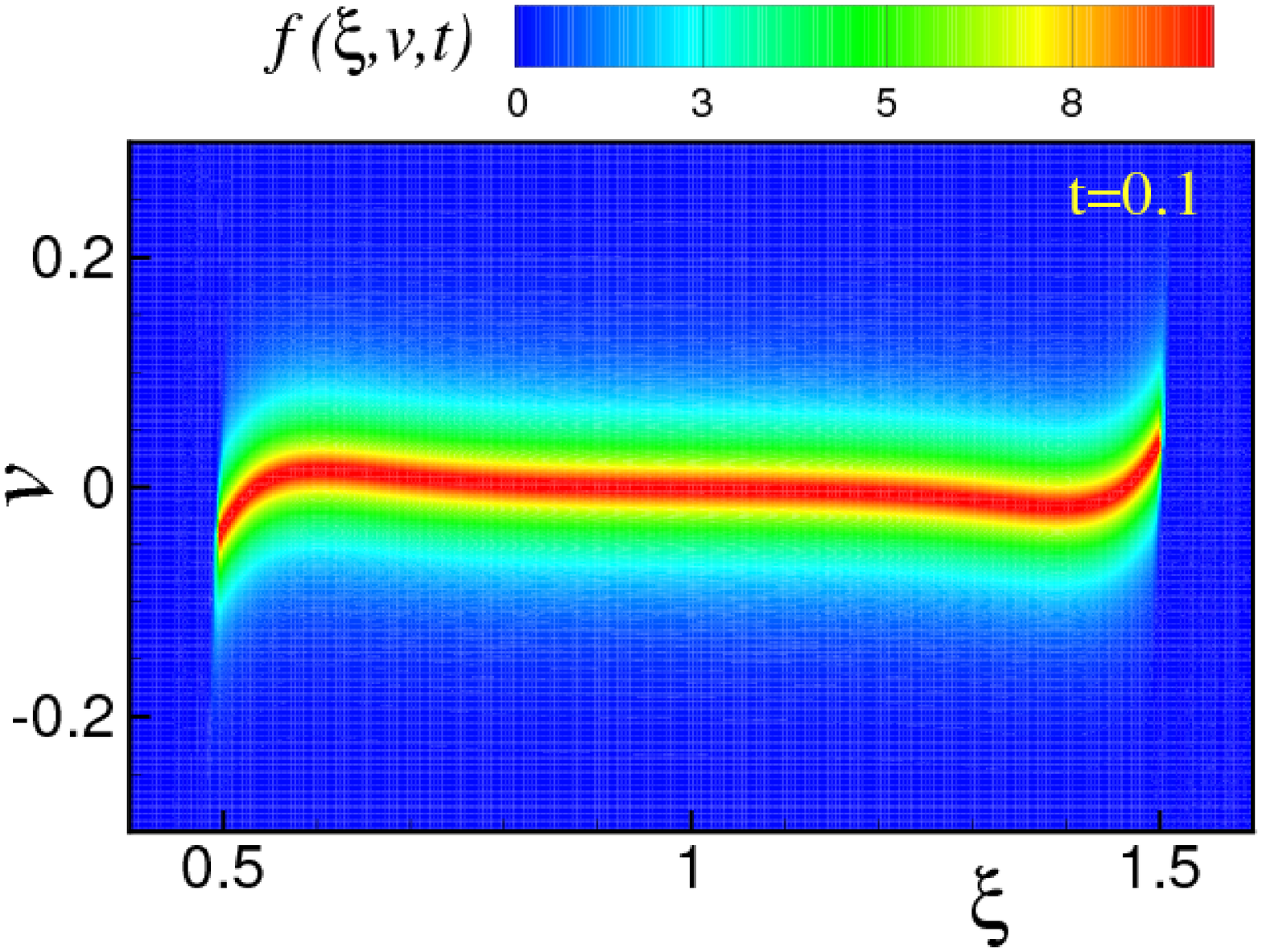} }  \hspace{-3mm}
                    \hbox{\includegraphics[width=0.4\textwidth]{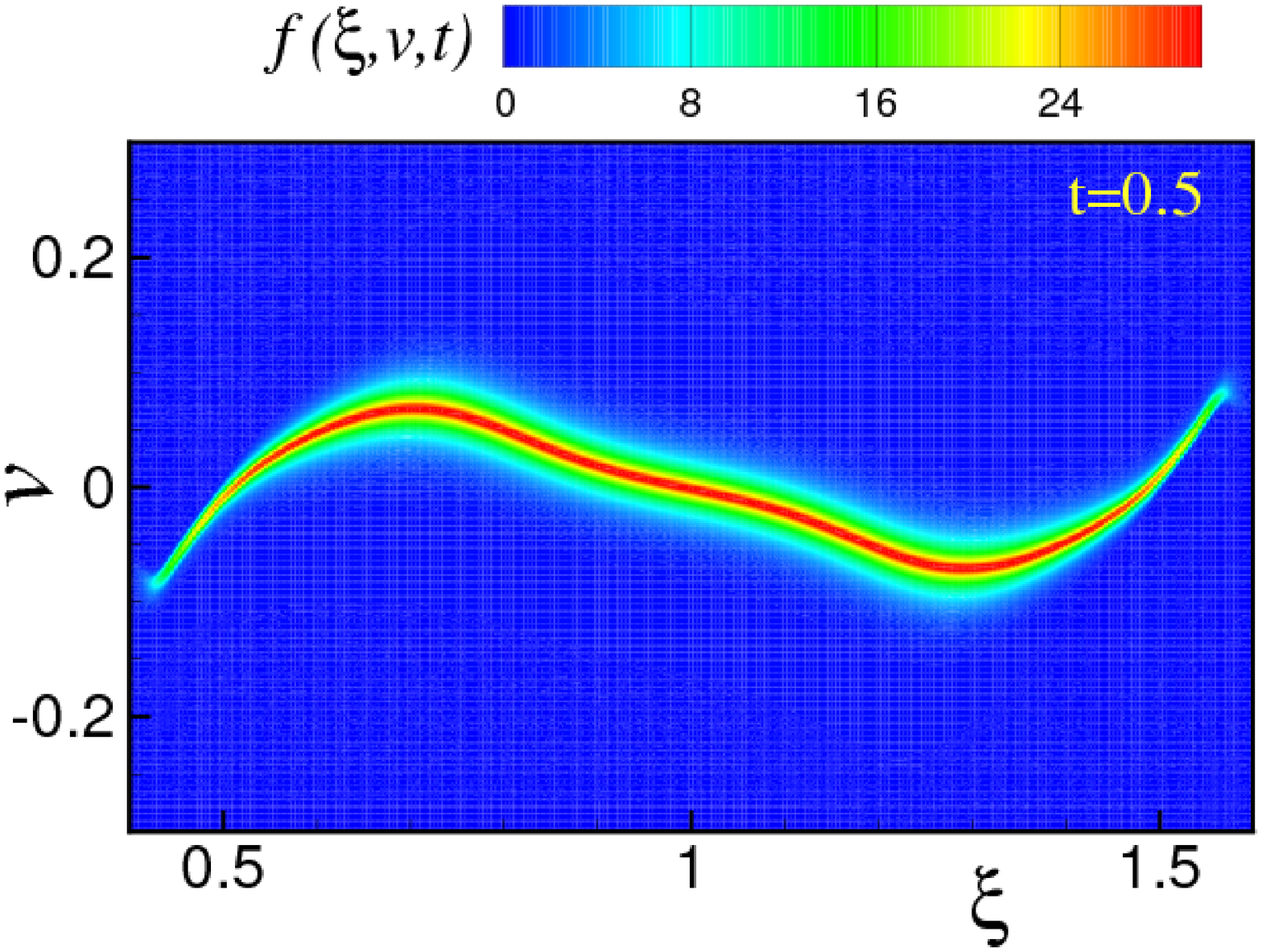} }     }
\centerline{\hbox{\includegraphics[width=0.4\textwidth]{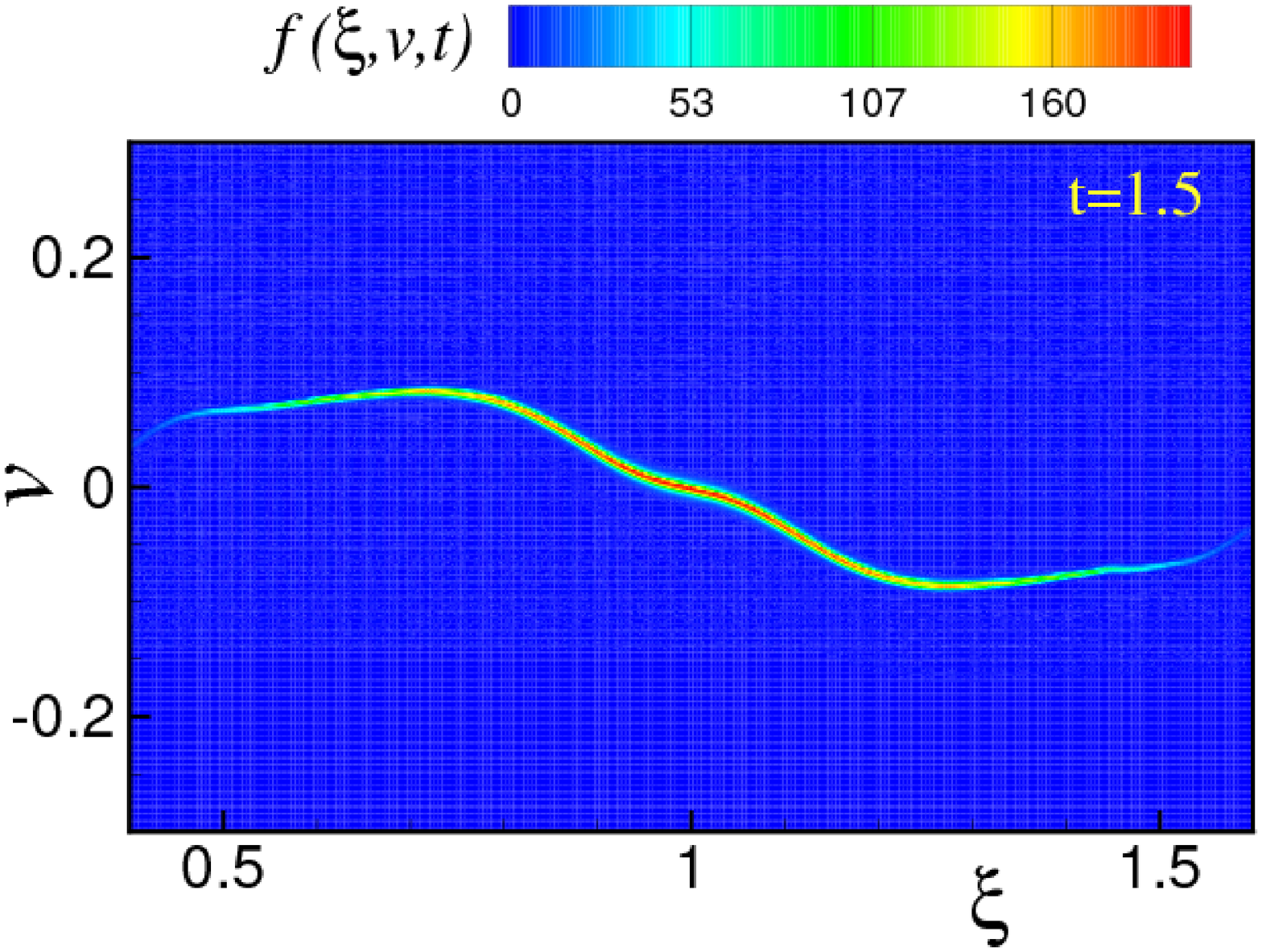} }  \hspace{-3mm}
                    \hbox{\includegraphics[width=0.4\textwidth]{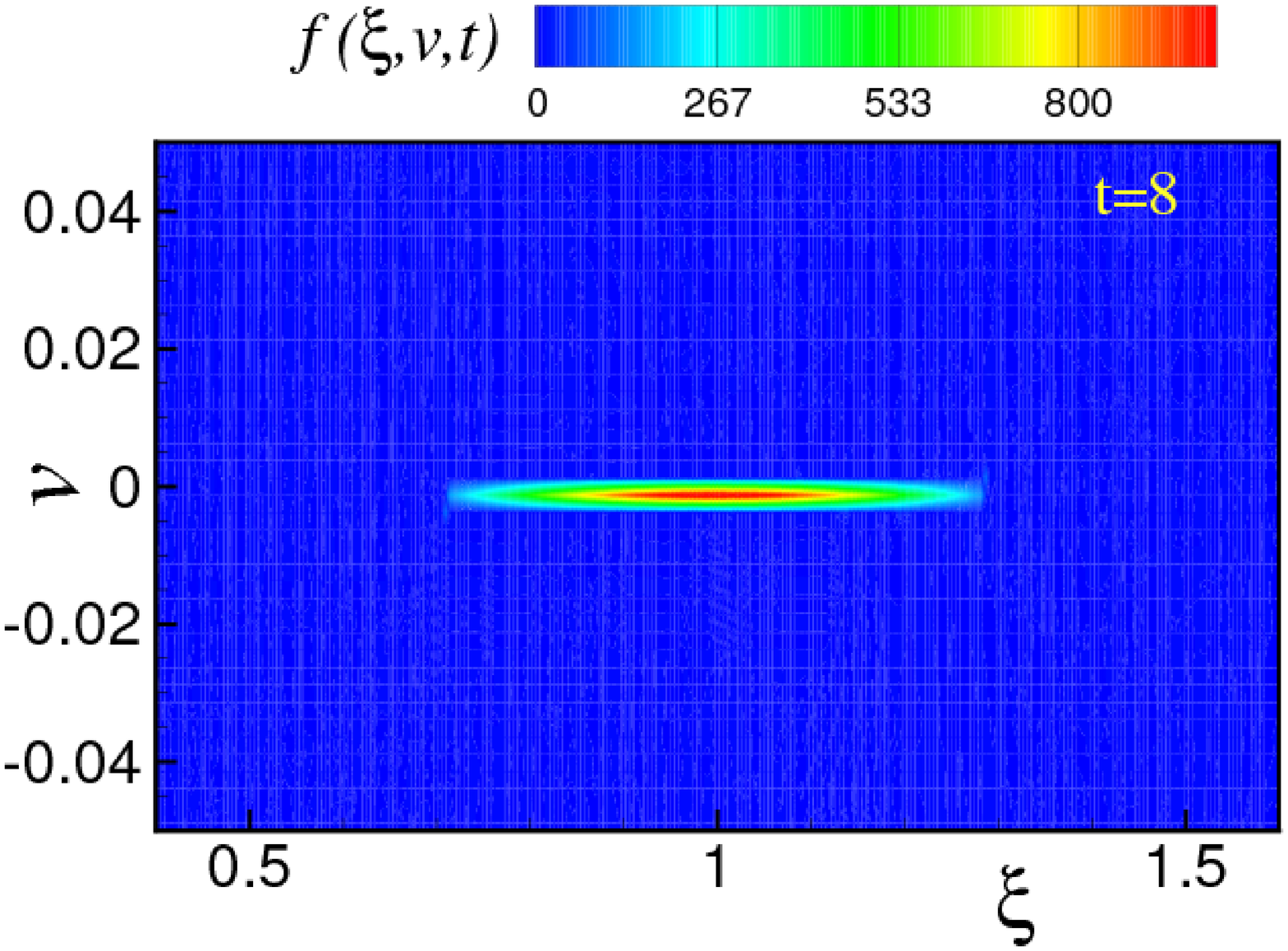} }     }                   
\caption{Evolution of distribution function $f(\xi,v,t)$ with $\xi=x-u_{\rm f} t$ in a catastrophic 
phase. The initial state of the system is given by equation (\ref{initial1}) where $X=2$. 
The parameters of the potential $\Phi$ are set to $c_1=1.5$, $c_2=1.0$, $d_1=0.05$ and $d_2=0.1$. 
Snapshots correspond to $t=0.1$, $0.5$, $1.5$, and $8$. Note the scale of $v$-axis in the bottom-right panel. 
The distribution function eventually converges to $f(\xi,v,t)=\rho(\xi)\delta(v-u_{\rm f})$.}
\label{fig1}
\end{figure*}

\begin{figure}
\centerline{\hbox{\includegraphics[width=0.34\textwidth]{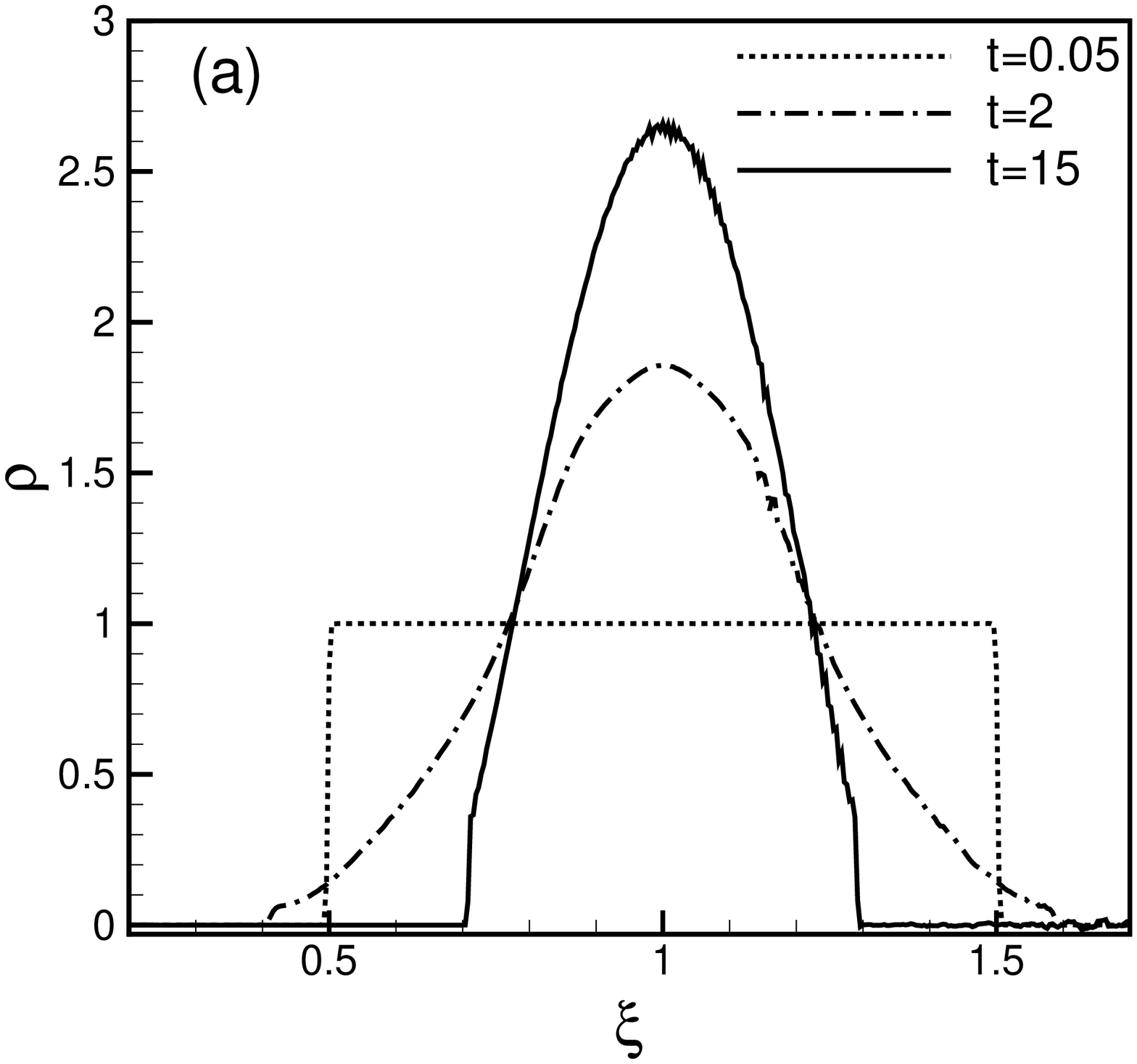} } \hspace{-5mm} 
                    \hbox{\includegraphics[width=0.34\textwidth]{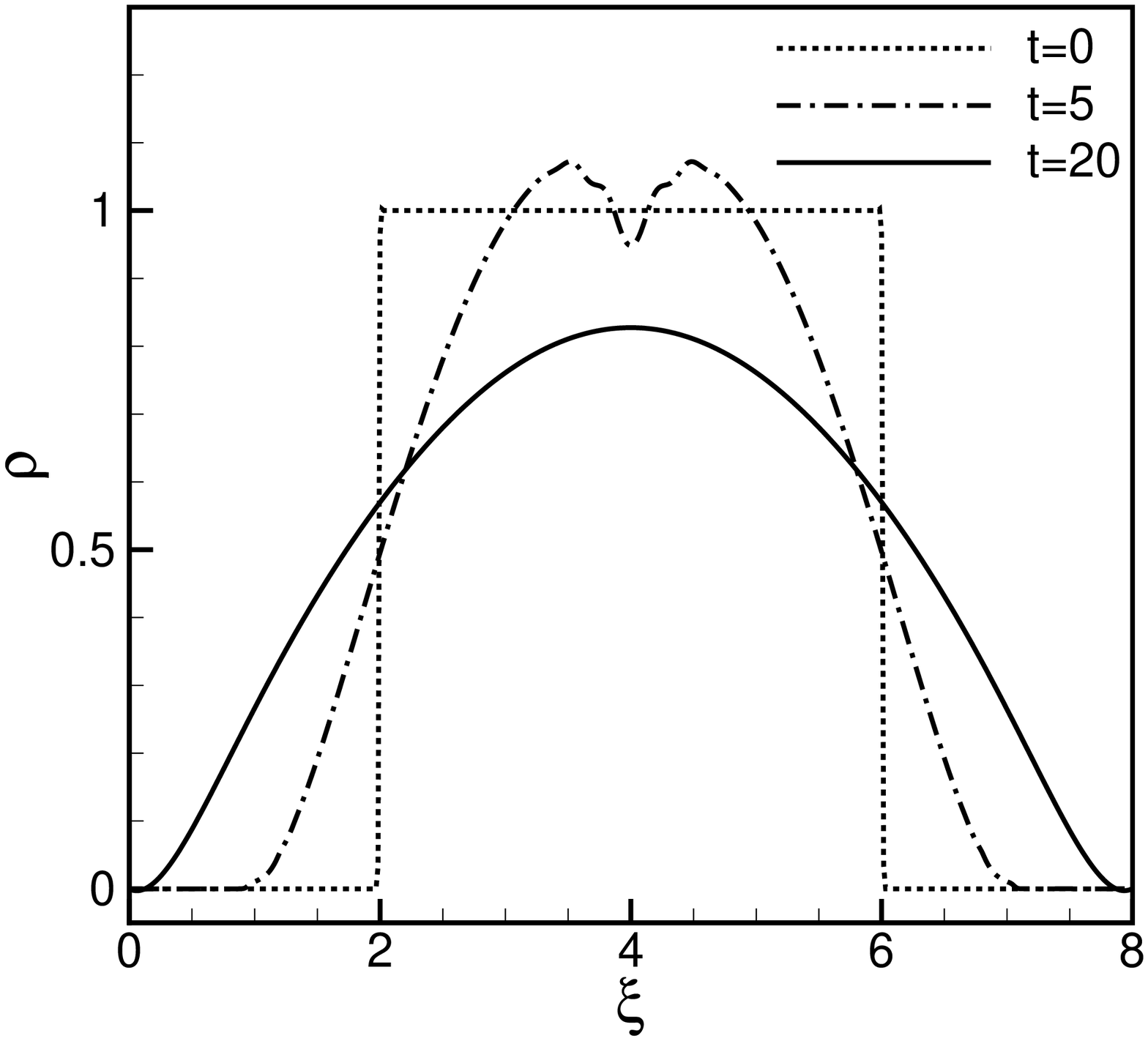} } \hspace{-5mm}
                    \hbox{\includegraphics[width=0.34\textwidth]{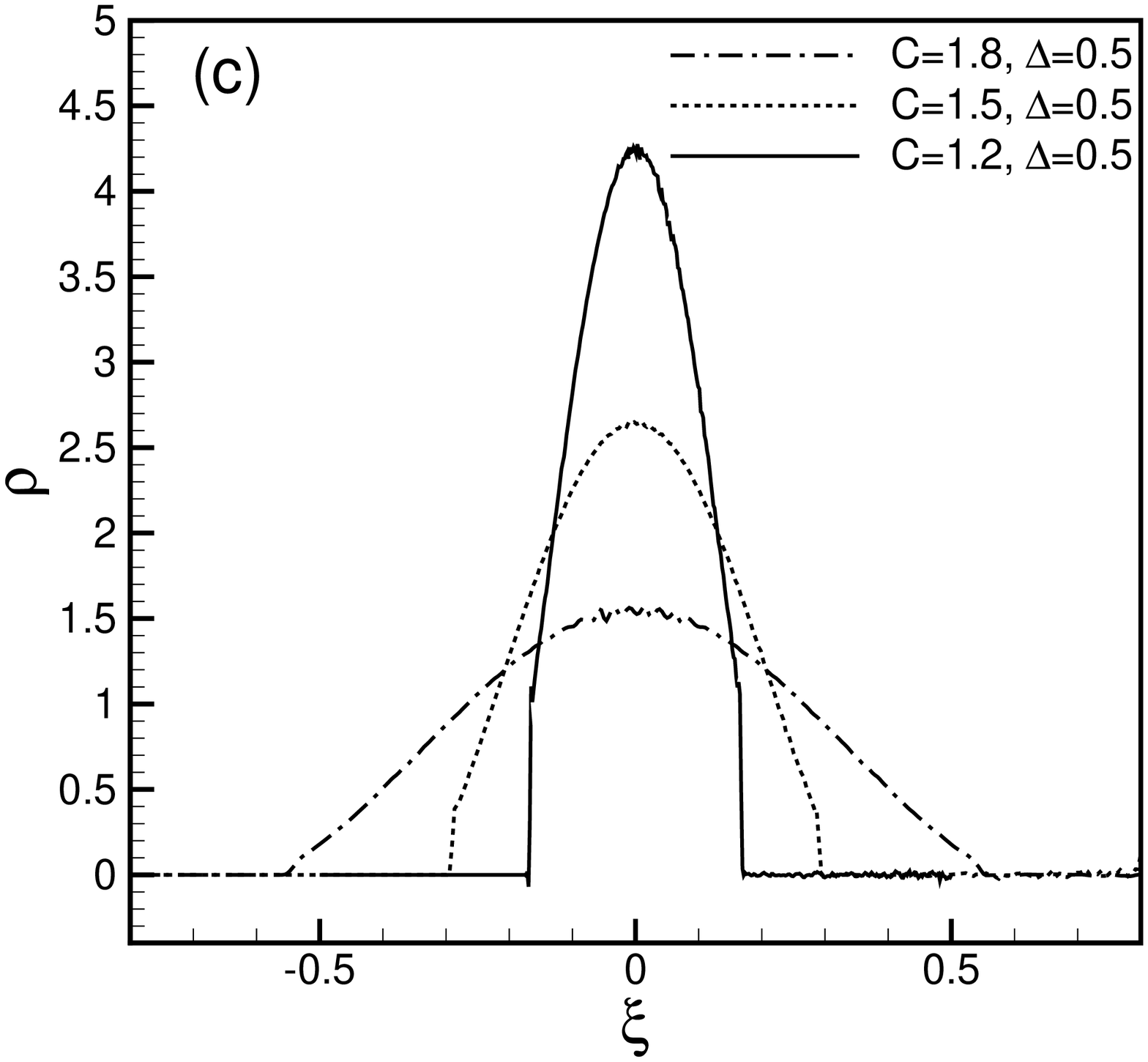} } 
                     }
\caption{(a) Evolution of spatial density $\rho$ obtained from the simulations of the Fokker-Planck 
equation in catastrophic phase for $\beta=2$, $c_1=1.5$, $c_2=1$, $d_1=0.05$, and $d_2=0.1$. 
The initial DF has been computed from equation (\ref{initial1}) with $X=2$. (b) Evolution of $\rho$ in $H$-stable 
phase for a model with $\beta=0.5$, $c_1=2.2$, $c_2=1$, $d_1=0.05$, and $d_2=0.1$. The density 
profile spreads to a finally flat state. The initial distribution is similar to equation (\ref{initial1}) but with 
$X=8$. (c) Relaxed stable density profiles for three models in the catastrophic phase. In all cases we 
have used $\beta=0.5$, $c_2=1$, $d_1=0.05$, and $d_2=0.1$. It is evident that the flock length 
increases proportional to $C$.}
\label{fig2}
\end{figure}

For long-range attractions and short-range repulsions, we define new parameters 
$C=c_1/c_2>1$ and $\Delta=d_1/d_2<1$, use the initial distribution function 
\begin{eqnarray}
f_0(x,v) = f(x,v,0)= \left \{
\begin{array}{lll}
10 \, e^{-20 \vert v \vert} & ,  & X/4<x<3 X/4, \\
0 & , &  x< X/4 ~ {\rm or} ~  x > 3X/4, 
\end{array}
\right.
\label{initial1}
\end{eqnarray}
and solve equation (\ref{eq:kinetic}) using the procedures of section \ref{sec:Numerical-methods}.
Figure \ref{fig1} shows the snapshots of $f(x,v,t)$ at four different times. At $t=8$, the DF $f$ 
very well approximates the steady catastrophic phase with $C \Delta<1$ and $X=2$. It is seen 
that the distribution function eventually takes a spiky form (resembling the Dirac delta function) 
in the velocity subspace, and all particles move with the same velocity of the carrier fluid. 
We have observed the spiky nature of the DF for $H$-stable phase as well. These results 
justify the assumption of mono-kinetic DFs used in the derivation of hydrodynamic equations.
Figure \ref{fig2}(a) demonstrates the spatial density $\rho(x,t)$ at three different times.
The steady state density profile is the one seen at $t=15$. The pattern of $f$ in the $x$-subspace 
depends on $C$ and $\Delta$. The drag coefficient $\beta$ and $f_0(x,v)$ control the pattern 
and speed of intermediary stages, and not the final state. For $C \Delta >1$, the initial DF 
evolves to an $H$-stable phase (see Figure \ref{fig2}(b)) whose density profile spreads in the 
$x$-domain while the average velocity of particles tends to fluid velocity. 

In two dimensional studies the curve $C \Delta^2=1$ is the separatrix (in the parameter space) 
between $H$-stable and catastrophic phases \cite{Ors06}, but in one dimensional systems the 
separatrix is the curve $C \Delta =1$ \cite{Lev09}, which is in agreement with our results. 
The final stable density pattern in the catastrophic phase correlates with the value of $C \Delta$ 
(Figure \ref{fig2}(c)). It is evident that increasing $C \Delta$ yields larger stable flock lengths. 
  
\subsection{Analytical solutions for steady flocks}
\label{sec:analytical}
The final state of $f$ in Figure \ref{fig1} suggests that we can find an exact solution for 
$\rho(x,t)$ in catastrophic conditions. Since particles eventually acquire the same velocity 
of the carrier fluid, one can start with the monokinetic DF $f(x,v,t)=\rho(x,t)\delta(v-u(x,t))$ 
and the traveling density pattern $\rho(x,t)=\rho(x-u_{\rm f}t)$ where $u(x,t)=u_{\rm f}$. 
Substituting these assumptions into the hydrodynamic equations (\ref{eq:hydro-mass}) 
and (\ref{eq:hydro-momentum}), and using the co-moving coordinate $\xi = x-u_{\rm f} t$, 
we obtain the integral equation
\begin{eqnarray}
\partial_{\xi} \Phi *\rho = \frac{c_1}{d_{\rm 1}}({\cal B}_1-{\cal A}_1)+\frac{c_2}{d_{\rm 2}}({\cal A}_2-{\cal B}_2)=0,
\label{eq:Usro}
\end{eqnarray}
where
\begin{eqnarray}
{\cal A}_i(\xi) = \int_0^\xi e^{-\frac{1}{d_i}(\xi-\eta) } \rho(\eta) \, \dif \eta,~~
{\cal B}_i(\xi) = \int_\xi^L e^{\frac{1}{d_i}(\xi-\eta) } \rho(\eta) \, \dif \eta,~~i=1,2,
\end{eqnarray}
and $L$ is a to-be-determined flock length. Differentiating equation (\ref{eq:Usro}) four times with 
respect to $\xi$, and eliminating ${\cal A}_i$ and ${\cal B}_i$ ($i=1,2$) from calculations, lead to 
\begin{eqnarray}
\label{eq:rho_stable}
\frac{\partial^3{\rho}}{\partial{\xi}^3}- \frac{ C\Delta - 1 }{ d_1 d_2 ( C - \Delta )} \frac{\partial \rho}{\partial \xi}=0,
\end{eqnarray}
whose solution is 
\begin{eqnarray}
\rho=b_0+b_1e^{\alpha_1 \xi}+b_2e^{\alpha_2 \xi} ,~~
\alpha_{1} =+\sqrt{ \frac{ C\Delta - 1 }{ d_1 d_2 ( C - \Delta )}},~~
\alpha_{2} =-\sqrt{ \frac{ C\Delta - 1 }{ d_1 d_2 ( C - \Delta )}},
\label{eq:rho}
\end{eqnarray}
Due to long-range attraction and short-range repulsion we always have $ C > \Delta $.
Therefore, $\alpha_i$ ($i=1,2$) will be real numbers if $C\Delta > 1$, and pure imaginary numbers 
otherwise. We substitute from (\ref{eq:rho}) into equation (\ref{eq:Usro}) and its first derivative 
with respect to $\xi$, then evaluate the four resulting equations at $\xi=0$ and $\xi=L$ to obtain 
\begin{figure}
\centerline{\hbox{\includegraphics[,width=0.45\textwidth]{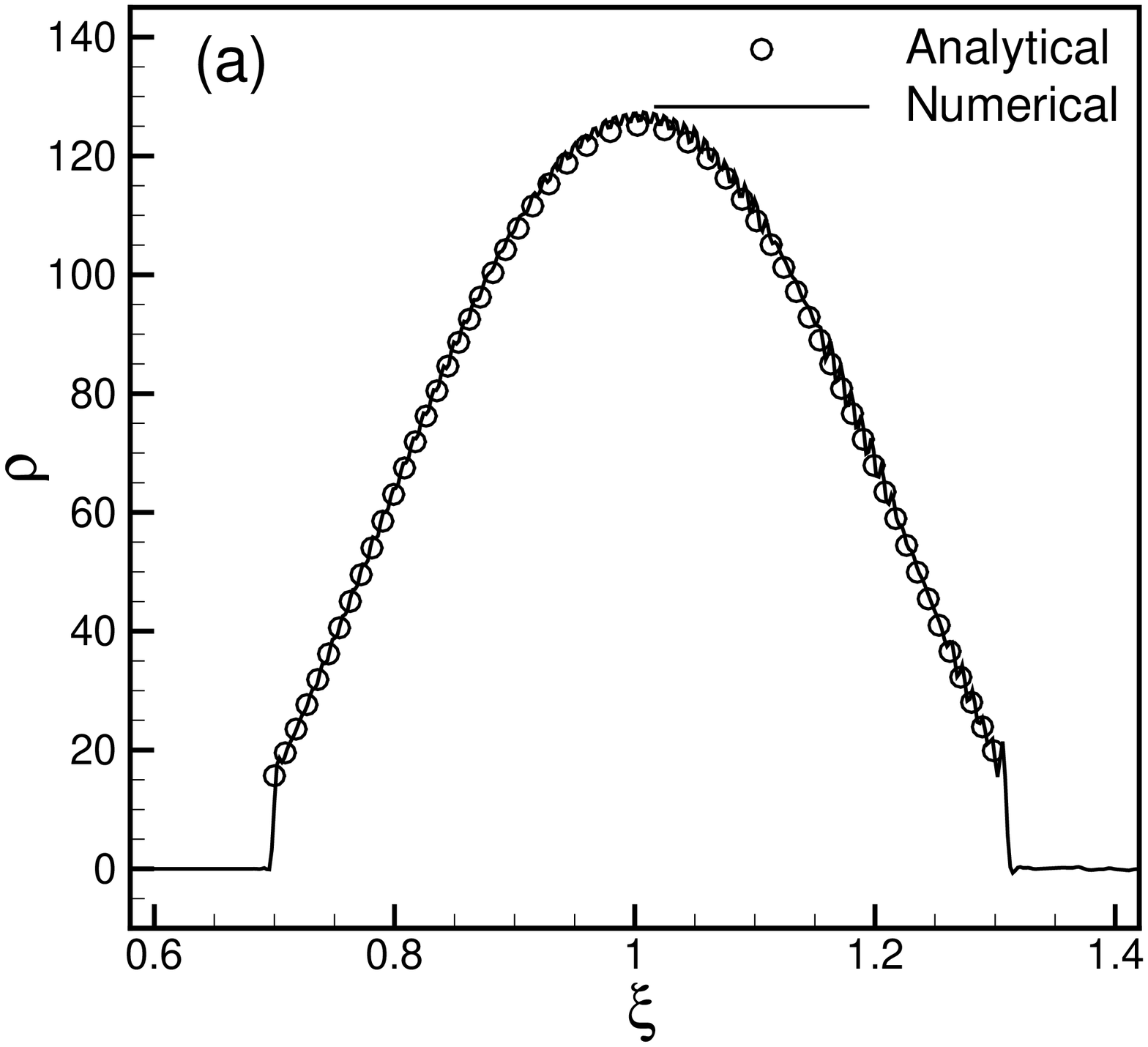} } 
                    \hbox{\includegraphics[width=0.45\textwidth]{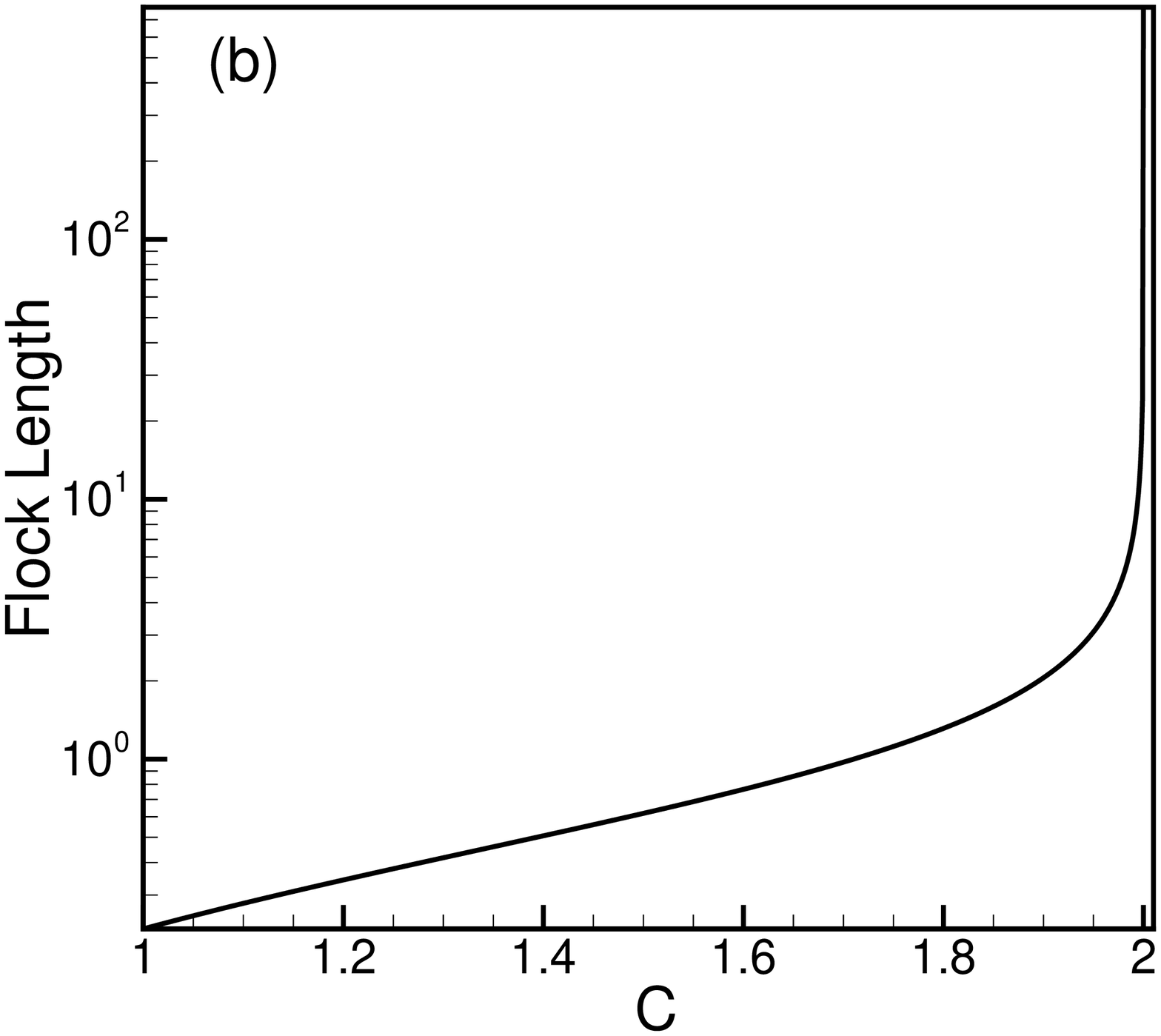} }    }
\caption{(a) Relaxed density $\rho$ in a catastrophic phase computed using the Fokker-Planck 
equation (solid line) and analytical results of section \ref{sec:analytical} (circles). Model
parameters are $d_1= 0.1$, $d_2=0.05$, $c_1=1.5$, and $c_2=1$. (b) Variation of the flock 
length $L$ in terms of $C$ for a model with $d_1=0.05$, $d_2=0.1$ and $C_2=1$.}
\label{fig3}
\end{figure} 
\begin{eqnarray}
\left [ 
\begin{array}{ccc}
 B_{11}  &  B_{12}                        & B_{13} \\
 B_{21}  &  B_{22}                        & B_{23} \\
 B_{11}  &  e^{\alpha_1 L} B_{13} & e^{\alpha_2 L} B_{12}  \\
 B_{21}  &  e^{\alpha_1 L} B_{23} & e^{\alpha_2 L} B_{22} 
 \end{array}
 \right ] \cdot 
 \left \{
\begin{array}{c}
b_0  \\
b_1  \\
b_2
\end{array}
\right \}
= {\bf 0},
\label{eq:determine-b-vector}
\end{eqnarray}
where 
\begin{eqnarray}
B_{i1}=d_i, ~~  B_{i2}=\frac{C^{i-2}(1-\alpha_1 d_i)(C-\Delta )}{ d_2 (1 - \Delta^2)}, ~~
B_{i3}=\frac{C^{i-2}(1-\alpha_2 d_i)(C - \Delta )}{d_2 (1  - \Delta ^2)},
\end{eqnarray}
and $i=1,2$. Non-trivial solutions exist for $b_0$, $b_1$ and $b_2$ should the determinants of 
all $3 \times 3$ sub-matrixes in the coefficient matrix of equation (\ref{eq:determine-b-vector}) vanish. 
This gives the unique solution:
\begin{eqnarray}
e^{\alpha_1 L}=\frac{ C (1-\Delta )[d_1 \alpha_1 (C-\Delta) + 1 -C\Delta ]}
{d_1 \alpha_1 (C - \Delta )(C \Delta + C -2)+(C\Delta -1)(C\Delta+C-2\Delta )},
\label{eq:expalpha}
\end{eqnarray}
from which one finds $L$. Equation (\ref{eq:rho}) shows that $\alpha_1$ has a real positive 
value for $H$-stable solutions with $C\Delta>1$. By substituting for $\alpha_1$ from (\ref{eq:rho}) 
into (\ref{eq:expalpha}), and imposing $C\Delta>1$, $C>1$ and $\Delta<1$, one can verify that 
$e^{\alpha_1 L} < 1$. This corresponds to a non-physical flock length $L<0$ as is expected:
the $H$-stable phase does not admit any characteristic/flock length. In the catastrophic 
phase with a pure imaginary $\alpha_1$, the right hand side of equation (\ref{eq:expalpha}) becomes 
a complex number as $\exp[{\rm i} \, \gamma(C,\Delta)]$ with ${\rm i}=\sqrt{-1}$ and $0 \le \gamma \le 2\pi$. 
Our calculations show that the physical flock length corresponding to $\rho(\xi) \ge 0$ becomes 
$L={\rm i} \gamma/\alpha_1$. Once $L$ is computed, the constant coefficients $b_0$, $b_1$ 
and $b_2$ can be calculated from (\ref{eq:determine-b-vector}). 

Figure \ref{fig3}(a) compares $\rho(\xi)$ profiles obtained from (\ref{eq:rho}) and numerical 
solutions of the Fokker-Planck equation. There is an impressive match between the two sets 
of results, providing a bench mark for our more complex computations in section \ref{subsec:stability-unsteady}. 
In Figure \ref{fig3}(b) we have plotted $L$ versus the parameter $C$ for $\Delta =0.5$. It is seen 
that the flock length tends to infinity as $C\Delta \rightarrow 1$. This is because $\alpha_1\rightarrow 0^+i$, 
which indicates the transition boundary from the catastrophic to $H$-stable phase. For $C\Delta =1$, 
equation (\ref{eq:rho_stable}) transforms to $\partial^3{\rho}/\partial{\xi}^3=0$ whose solution is 
$\rho=b_0+b_1\xi+b_2 \xi^2$. Employing the boundary conditions at $\xi=0$ and $\xi=L$, one 
obtains a system of determinantal equations that have no root for finite values of $L>0$. We thus 
conclude that the boundary line $C\Delta =1$ in the parameter space belongs to the $H$-stable 
phase with no characteristic/flock length.

\section{Collective dynamics in unsteady flows}
\label{sec:unsteady-flows}

We now consider time-varying $u_{\rm f}(t)$, search for base-state solutions in the catastrophic 
phase, and investigate their stability. We think of a co-moving coordinate system with the velocity 
$u_{\beta}(t)$, and carry out the following change of spatial variable:
\begin{eqnarray}
\xi = x - \int_0^t {u_{\beta}(\tau)} \, d\tau.
\label{eq:traveling-coordinates}
\end{eqnarray}
The velocity $u_{\beta}(t)$ is yet to be determined with the aim of eliminating 
explicit time-dependent terms from the continuity and momentum equations. In the co-moving 
frame, the density and streaming velocity of particles are expressed as $\tilde\rho(\xi,t)=\rho(x(\xi,t),t)$ 
and $\tilde u(\xi,t)=u(x(\xi,t),t)$, respectively. Equations (\ref{eq:hydrodynamic}) are thus 
transformed to
\begin{eqnarray}
\frac{\partial{\tilde{\rho}}}{\partial{t}} &+& (\tilde{u}-u_{\beta})  \frac{\partial \tilde{\rho}}{\partial \xi}
+\tilde{\rho} \frac{\partial \tilde u}{\partial \xi} = 0, \label{eq:continumeM1} \\
\frac{\partial{\tilde{u}}}{\partial{t}} &+& (\tilde{u}-u_{\beta}) \frac{\partial \tilde u}{\partial \xi}+
\beta[\tilde{u}-u_{\rm f}(t)]+\partial_{\xi} \Phi *\tilde \rho = 0, \label{eq:continumeM2}
\end{eqnarray}
which explicitly depend on $t$ through the terms involving $u_{\beta}(t)$ and $u_{\rm f}(t)$.
These terms are eliminated should $u_{\beta}$ be the particular solution of the following 
ordinary differential equation
\begin{eqnarray}
\frac{\dif u_{\beta}}{\dif t}+\beta[u_{\beta}(t)-u_{\rm f}(t)]=0.
\label{eq:pulse}
\end{eqnarray}
Consequently, equations (\ref{eq:continumeM1}) and (\ref{eq:continumeM2}) become  
\begin{eqnarray}
\partial \tilde \rho_0 /\partial t=0,~~ \tilde u_0=u_{\beta},~~  \partial_{\xi} \Phi * \tilde \rho_0 = 0,
\label{eq:base-state}
\end{eqnarray}
where $\tilde \rho_0$ and $\tilde u_0$ are {\it traveling wave solutions}. Therefore, while the 
invariant shape of the spatial density profile given in equation (\ref{eq:rho}) travels according to 
(\ref{eq:traveling-coordinates}), the streaming velocity of particles varies over time but it is not 
identical to $u_{\rm f}(t)$. Using (\ref{eq:pulse}), one can determine the effect of amplitude and 
frequency of a pulsating carrier fluid (similar to blood flow) on particle streaming. For instance, 
a periodic excitation $u_{\rm f}=A_0+A_1\sin(\Omega{t})$ yields
\begin{eqnarray} 
\tilde u = A_0 + \frac{\beta^2\,A_1}{(\Omega^2+\beta^2)}\sin(\Omega{t} ) -
\frac{\Omega{\beta}\,A_1}{(\Omega^2+\beta^2)}\cos(\Omega{t} ).  
\end{eqnarray}
This shows that increasing the pulse frequency suppresses the periodic nature of particle 
streaming: for $\Omega \gg 1$ particles will not follow the carrier fluid, unless $\beta \rightarrow \infty$. 
Our numerical simulations confirm this effect.

\subsection{Stability of unsteady flocks}
\label{subsec:stability-unsteady}

An important question is whether the collective flock of particles is stable under time-varying fluid 
velocities. To answer this question, we perturb macroscopic quantities as 
$\tilde u = \tilde u_0(t)+ \tilde u_1(\xi,t)$ and $\tilde \rho = \tilde \rho_0(\xi)+ \tilde \rho_1(\xi,t)$, 
and linearize equations (\ref{eq:continumeM1}) and (\ref{eq:continumeM2}) to obtain
\begin{eqnarray}
\frac{\partial \tilde \rho_1}{\partial t} + \frac{\partial}{\partial \xi} \left ( \tilde u_1 \tilde \rho_0 \right ) = 0, ~~
\frac{\partial \tilde u_1 }{\partial t } + \beta \tilde u_1 + \partial_{\xi} \Phi * \tilde \rho_1 = 0.
\label{eq:perturbed}
\end{eqnarray}
These equations can be solved through a Fourier transformation in the time domain 
followed by an expansion in terms of suitable basis functions in the $\xi$-space. 
Nevertheless, equations (\ref{eq:perturbed}) are linear and solutions in the Fourier 
transform space will be decoupled for different frequencies. We can thus assume 
$\tilde u_1 = U(\xi) e^{\Omega t}$ and $\tilde \rho_1=R(\xi) e^{\Omega t}$, substitute 
them in (\ref{eq:perturbed}) and eliminate $R(\xi)$ between the linearized continuity 
and momentum equations. This leaves us with the eigenvalue problem 
\begin{equation}
\int_0^L \partial_{\xi} \Phi(\vert \xi-\eta \vert ) \frac {\partial}{\partial \eta}
\left [ U(\eta) \tilde \rho_0(\eta) \right ] \,d\eta = \lambda U(\xi), ~~ \lambda = -\Omega(\Omega+\beta),
\label{eq:integral-equation}
\end{equation}
whose solutions can then be superposed in the $\lambda$-space (using a discrete summation
or a continuous integral, whichever applies) to obtain the most general expressions for $\tilde u_1$ 
and $\tilde \rho_1$. Integrating (\ref{eq:integral-equation}) by parts gives   
\begin{eqnarray}
\left [ 2 \partial_{\xi} \Phi(0^+) \, \tilde \rho_0(\xi) +\lambda \right ] U(\xi)  &+&  \partial_{\xi} \Phi(\vert \xi-L \vert ) \, \tilde \rho_0(L) U(L)
- \partial_{\xi} \Phi (\xi) \, \tilde \rho_0(0) U(0)   \nonumber \\ 
&+&  \int_0^L K(\xi,\eta) \, U(\eta) \, \dif \eta = 0,
\label{eq:fredholm-integral}
\end{eqnarray}
which is a Fredholm-type integral equation with the kernel 
$K(\xi,\eta) = \tilde \rho_0(\eta) \partial_{\xi \xi}\Phi (\vert \xi-\eta \vert )$. We could 
not find a complete set of basis functions to expand $U(\xi)$, and adopted an implicit 
numerical method to solve (\ref{eq:fredholm-integral}) and compute the eigenvalue 
$\lambda$. We divide the $\xi$-space to $N$ equally spaced intervals with the 
increments $\Delta \xi=L/N$, and take $U_j=U(j\Delta \xi)$ $(j=0,1,\ldots,N)$ as 
unknown variables. The definite integral involving the kernel is then computed using 
the trapezoidal rule and equation (\ref{eq:fredholm-integral}) is transformed to the 
linear eigensystem $\Amat \cdot \Uvec = \lambda \Uvec$ where $\Amat$ is a 
constant matrix and $\Uvec$ is a column vector assembled from $U_j$. The 
eigenvalue problem is solved using standard LAPACK libraries.

\begin{figure}
\centerline{\hbox{\includegraphics[,width=0.48\textwidth]{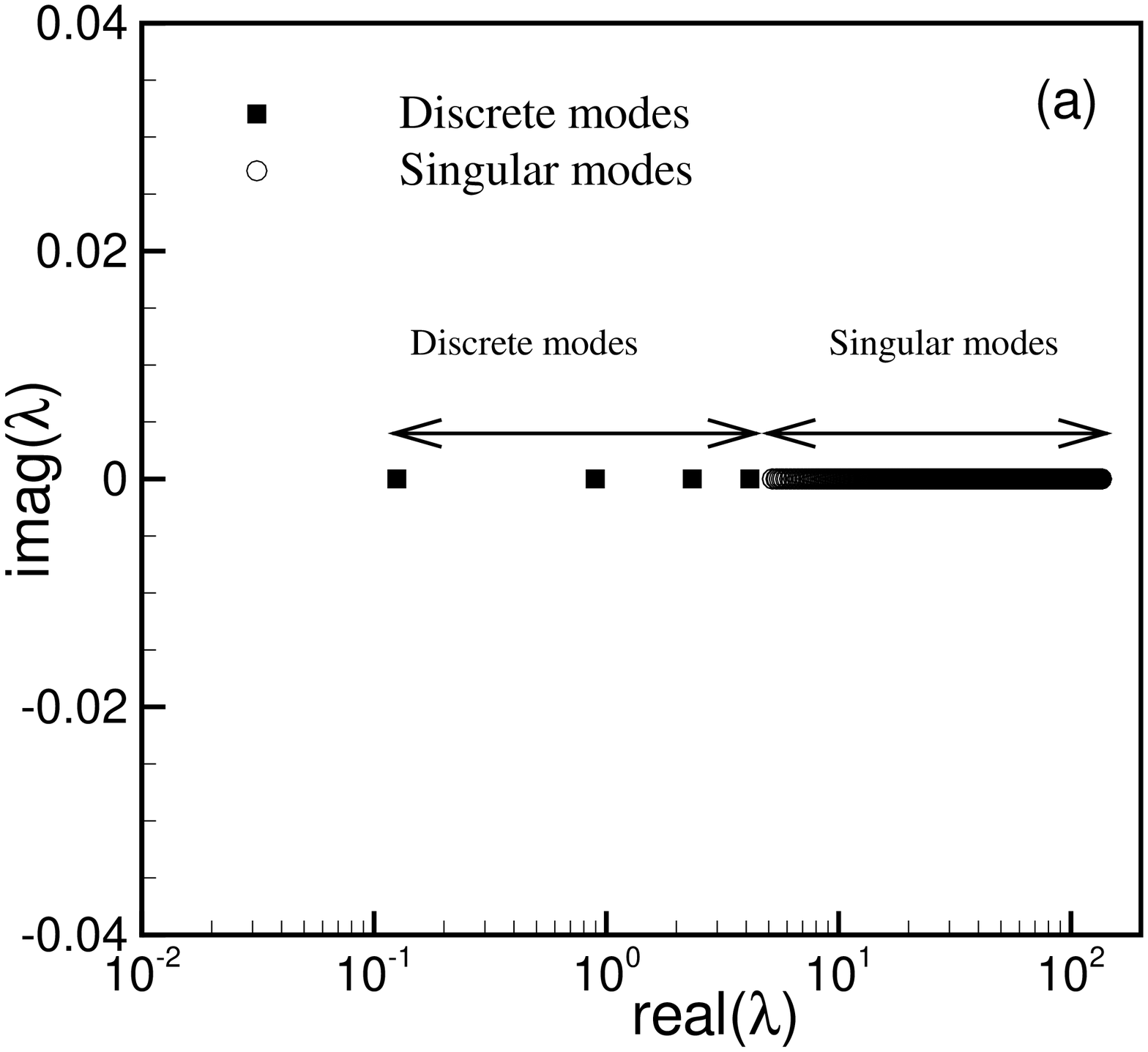} }  \hspace{-7mm}
                    \hbox{\includegraphics[width=0.48\textwidth]{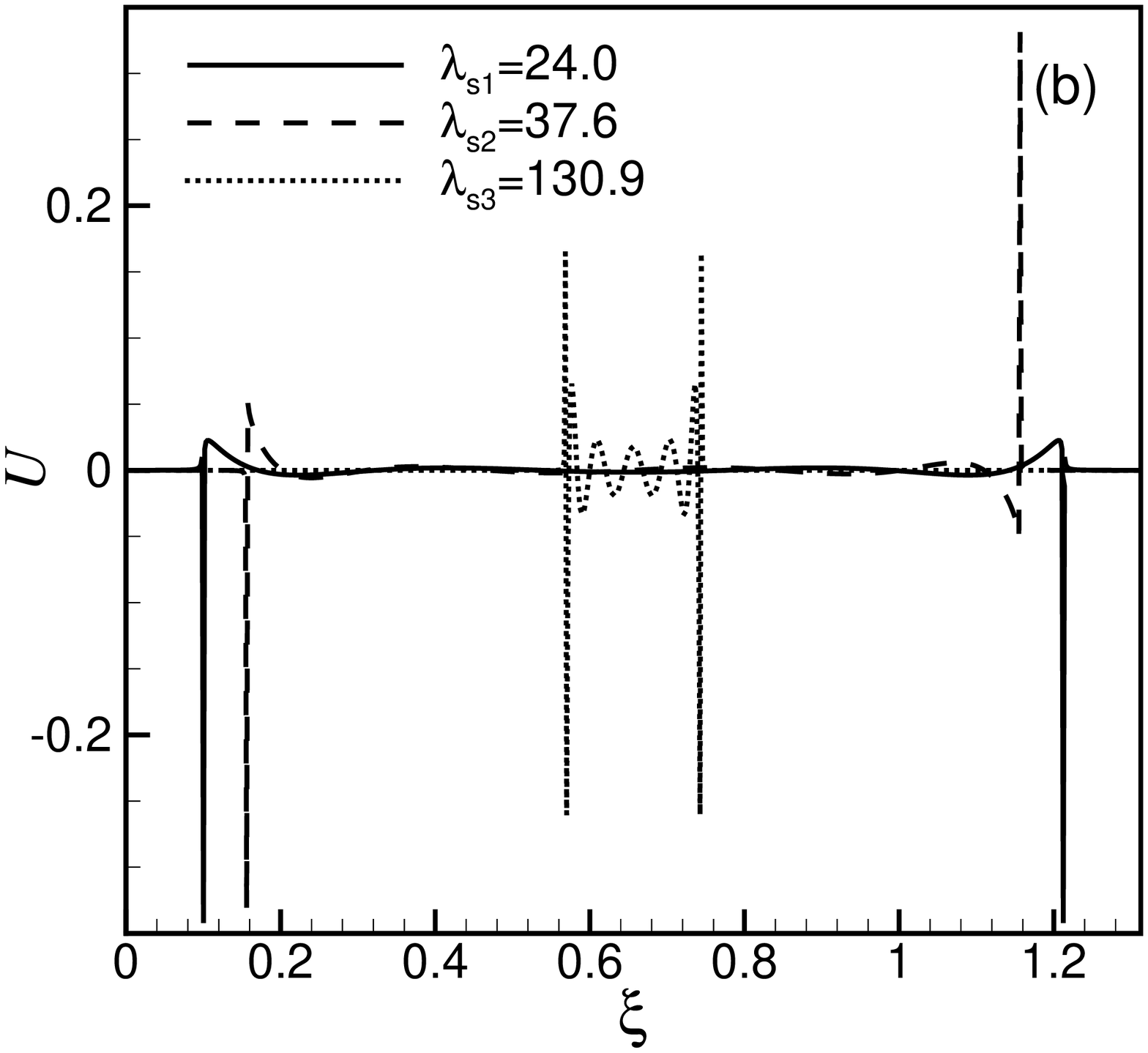} }   \hspace{-7mm}  }
\centerline{\hbox{\includegraphics[,width=0.48\textwidth]{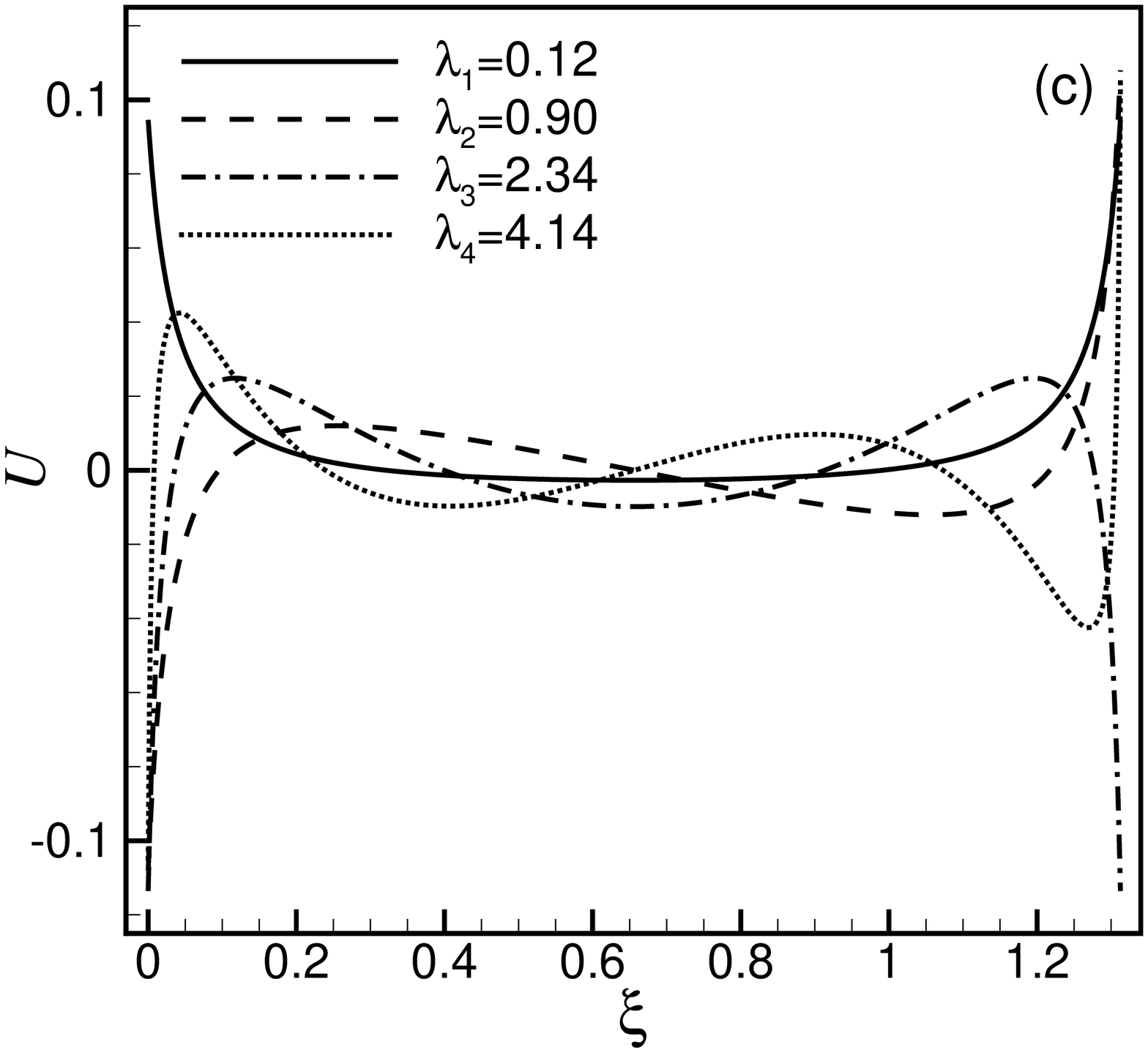} }  \hspace{-7mm}
                    \hbox{\includegraphics[width=0.48\textwidth]{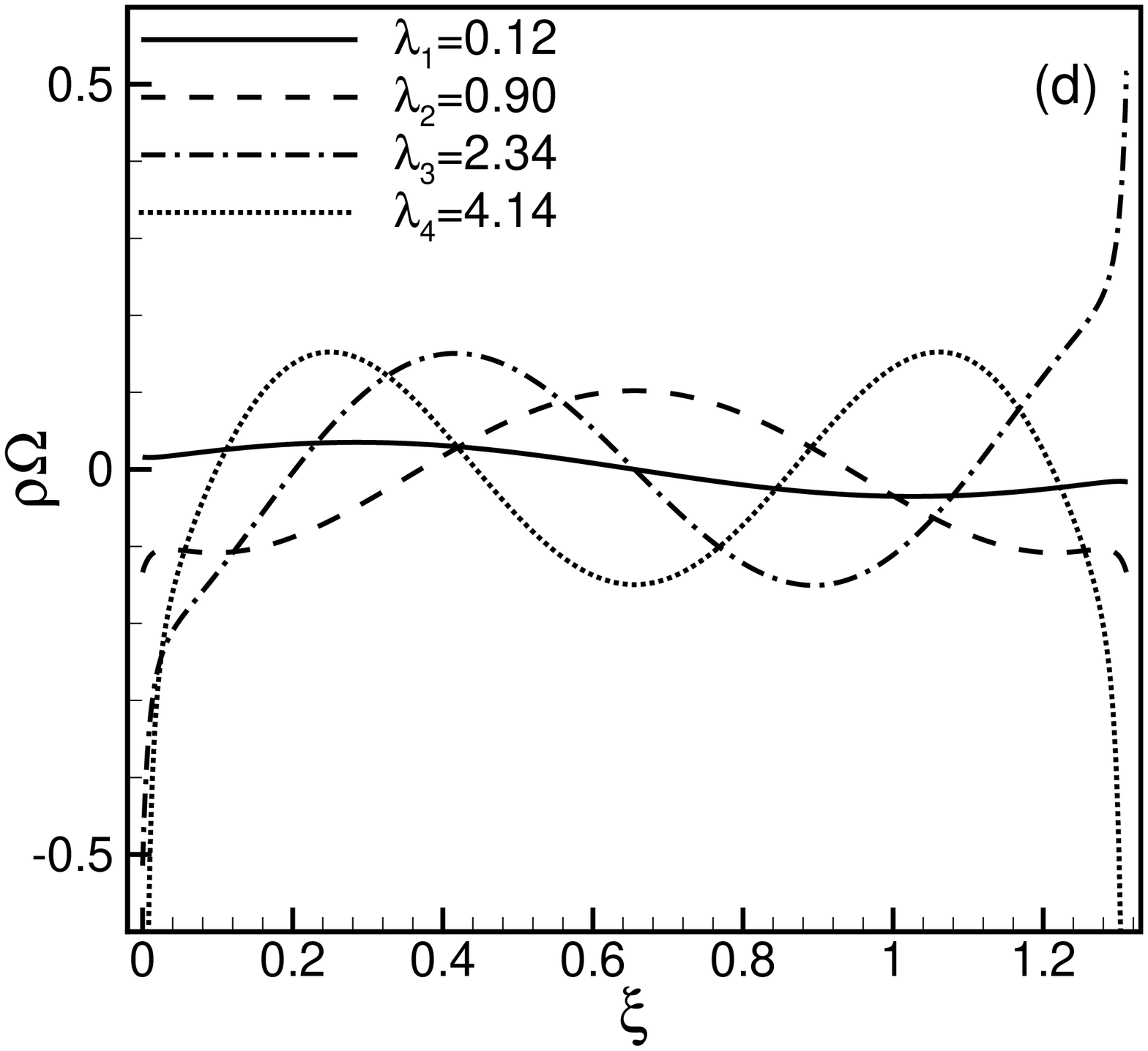} }   \hspace{-7mm}  }                    
\caption{(a) Eigenvalue spectrum of linearized equations (\ref{eq:perturbed}) in the vicinity 
of a time-varying $u_{\rm f}(t)$ for a model with $c_1=3.6$, $c_2=2$, $d_1=0.05$, and $d_2=0.1$. 
Equations (\ref{eq:perturbed}) are normalized with respect to the total number of particles
so that $\int_{0}^{L} \rho_0(\xi) \, \dif \xi=1$. Discrete global modes are shown with filled squares. 
Singular modes (circles) form a continuous family. Discrete modes converge within $0.5\%$ 
relative error by taking $N>400$ where $N$ is the number of discrete grid points in the $\xi$-space.
Increasing $N$ adds new singular modes to the continuous part of the spectrum. 
(b) Patterns of three singular modes for $\lambda_{s1}=24.0$, $\lambda_{s2}=37.6$ and $\lambda_{s3}=130.9$. 
One can verify that the locations of spikes are consistent with the roots of equation (\ref{eq:singular-eigenvalues}). 
(c) Perturbed velocity patterns of all discrete global modes. (d) Perturbed density patterns of all global modes.}
\label{fig4}
\end{figure} 

Interestingly, we find that the eigenspectrum contains a continuous spectrum of singular 
modes and a finite number of discrete global modes as shown in Figure \ref{fig4}(a). 
Singular modes emerge when the coefficient of $U(\xi)$ in the first term of 
(\ref{eq:fredholm-integral}) vanishes at $\xi_s \in [0,L]$ so that 
\begin{eqnarray}
\lambda_s = -2 \partial_{\xi} \Phi(0^+) \, \tilde \rho_0(\xi_s).
\label{eq:singular-eigenvalues}
\end{eqnarray}
In the Catastrophic phase, $\tilde \rho_0$ is a single-peaked function and equation 
(\ref{eq:singular-eigenvalues}) is satisfied at two points, $\xi_{s1}$ and $\xi_{s2}$. 
The mode shape $U(\lambda_s,\xi)$ corresponding to $\lambda_s$ thus becomes 
spiky at $\xi=\xi_{s1}$ and $\xi=\xi_{s2}$, and manages to annul the definite integral in 
(\ref{eq:fredholm-integral}). Singular modes have zero amplitudes at the boundaries,
so that  $U(\lambda_s,0)=U(\lambda_s,L)=0$, unless for the eigenvalue associated 
with $\xi_{s1} =0$ and $\xi_{s2}=L$. Figure \ref{fig4}(b) demonstrates the shapes 
of three singular modes. Discrete global modes have been shown in Figure \ref{fig4}(c).
It is evident that they are smooth and the number of their nodes increases proportional 
to $\lambda$. All eigenvalues that we find are positive real numbers and the mode 
frequencies are computed as $\Omega_{1,2}=-\beta/2\pm \sqrt{\beta^2/4-\lambda}$, 
which always has a negative real part. The traveling catastrophic solutions are therefore 
stable. Discrete modes will generate global, long-lived oscillatory patterns if $\beta \ll 1$ 
and $\beta^2 < 4\lambda$. A mode will be overdamped and critically damped for 
$\beta^2>4\lambda$ and $\beta^2=4\lambda$, respectively. Our computations show 
that $\lambda$ decreases by increasing $C$ and $\Delta$. For each eigenmode 
$U(\lambda,\xi)$ we obtain two {\em eigendensities}: 
\begin{eqnarray}
R_i(\lambda,\xi)=-\frac{1}{\Omega_i}\frac{\partial}{\partial \xi} \left [ 
\tilde \rho_0(\xi) U(\lambda,\xi) \right ], ~~ i=1,2.
\end{eqnarray}
Therefore, any perturbed state will be represented as a linear combination
\begin{eqnarray}
(\tilde \rho_1,\tilde u_1)=(a_1 \tilde \rho_{11}+a_2 \tilde \rho_{12} , a_1 \tilde u_{11}+a_2 \tilde u_{12}),
\label{eq:initial-DF-stability}
\end{eqnarray}
where $a_1$ and $a_2$ are constant coefficients and
\begin{eqnarray}
(\tilde \rho_{1j},\tilde u_{1j})=\left [ R_j(\lambda,\xi),U(\lambda,\xi) \right ] e^{\Omega_j t}.
\end{eqnarray}

\begin{figure}
\centerline{\hbox{\includegraphics[,width=0.34\textwidth]{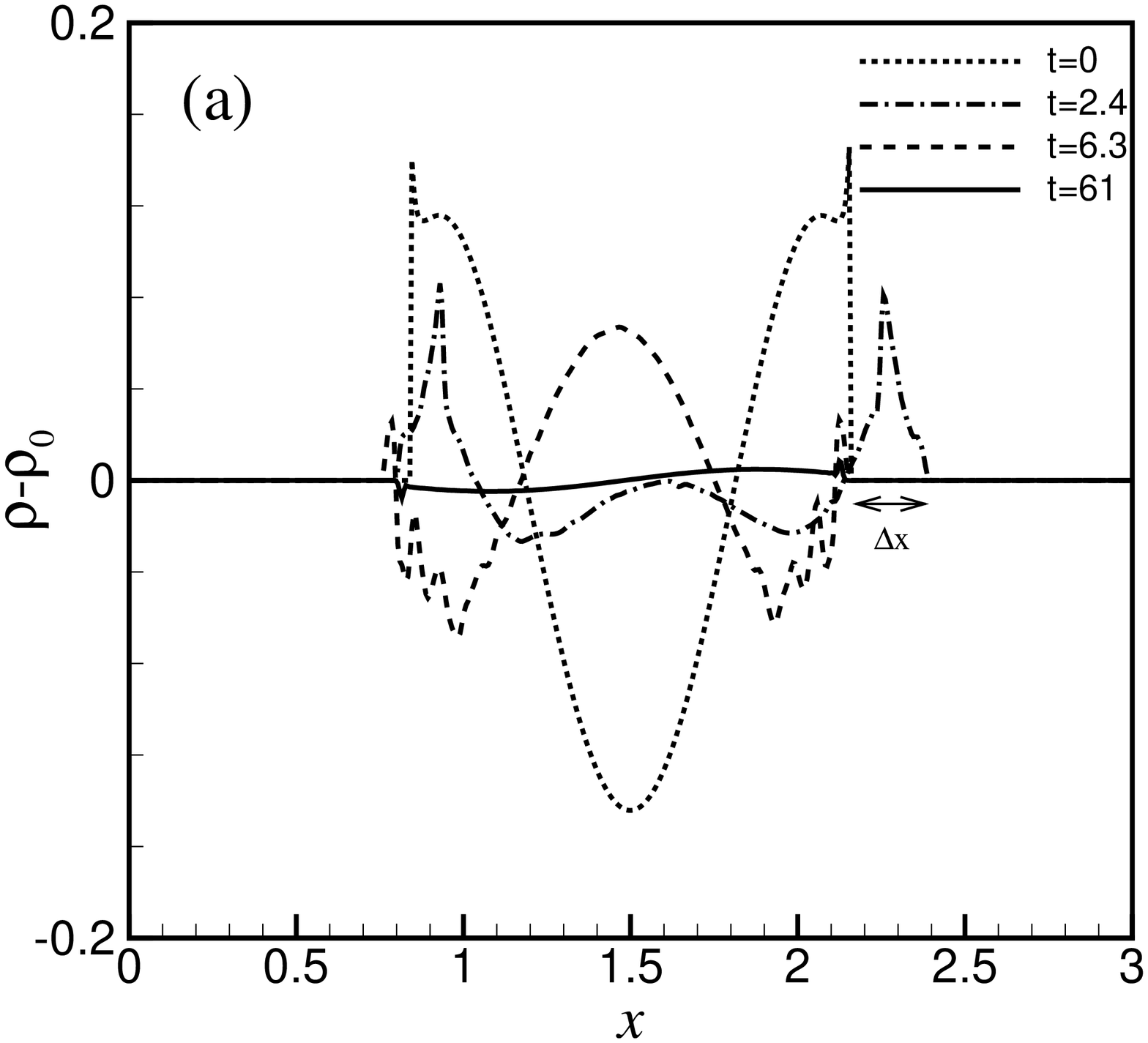} }  \hspace{-5mm}
                    \hbox{\includegraphics[width=0.34\textwidth]{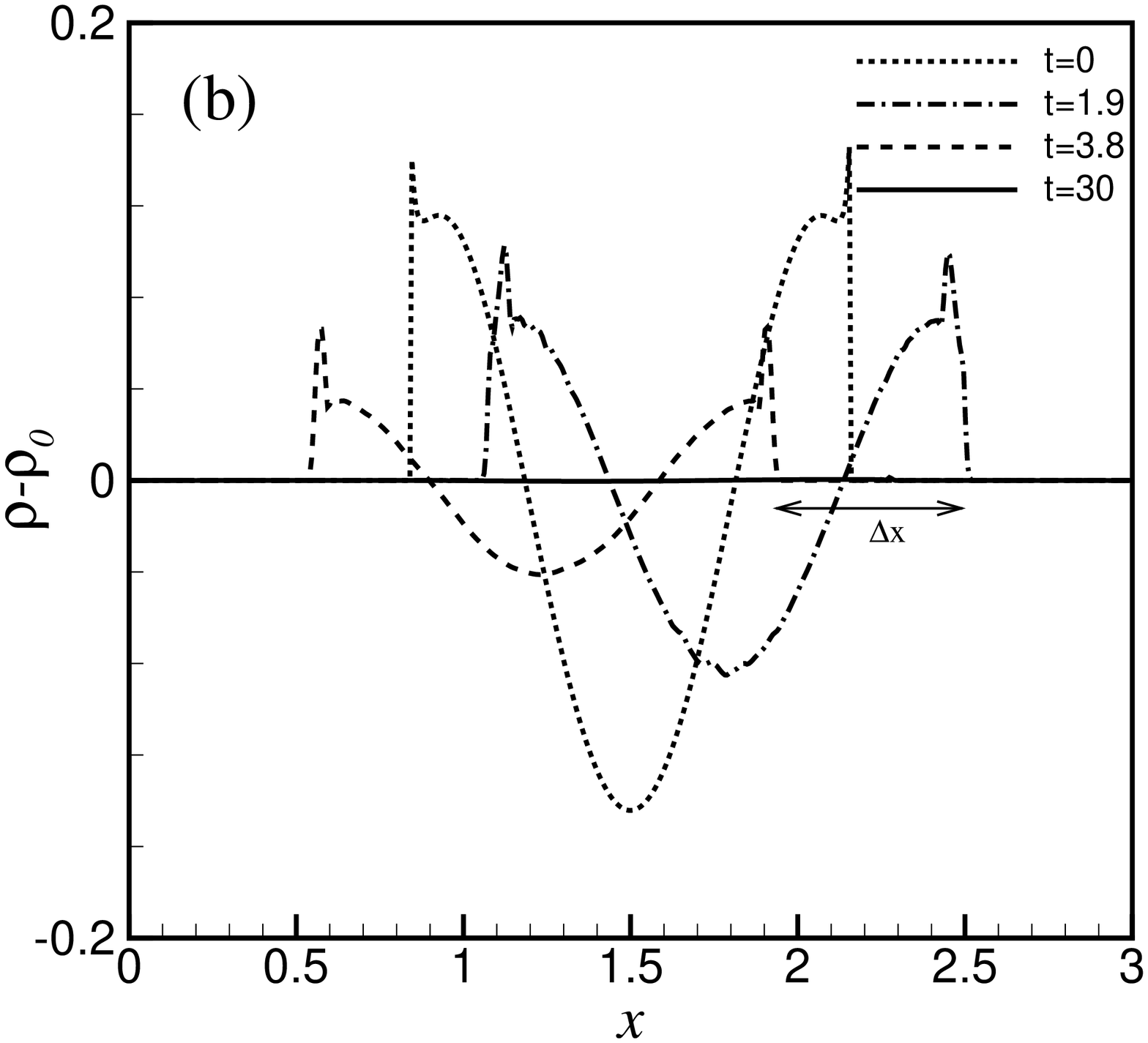} }   \hspace{-5mm} 
                    \hbox{\includegraphics[width=0.34\textwidth]{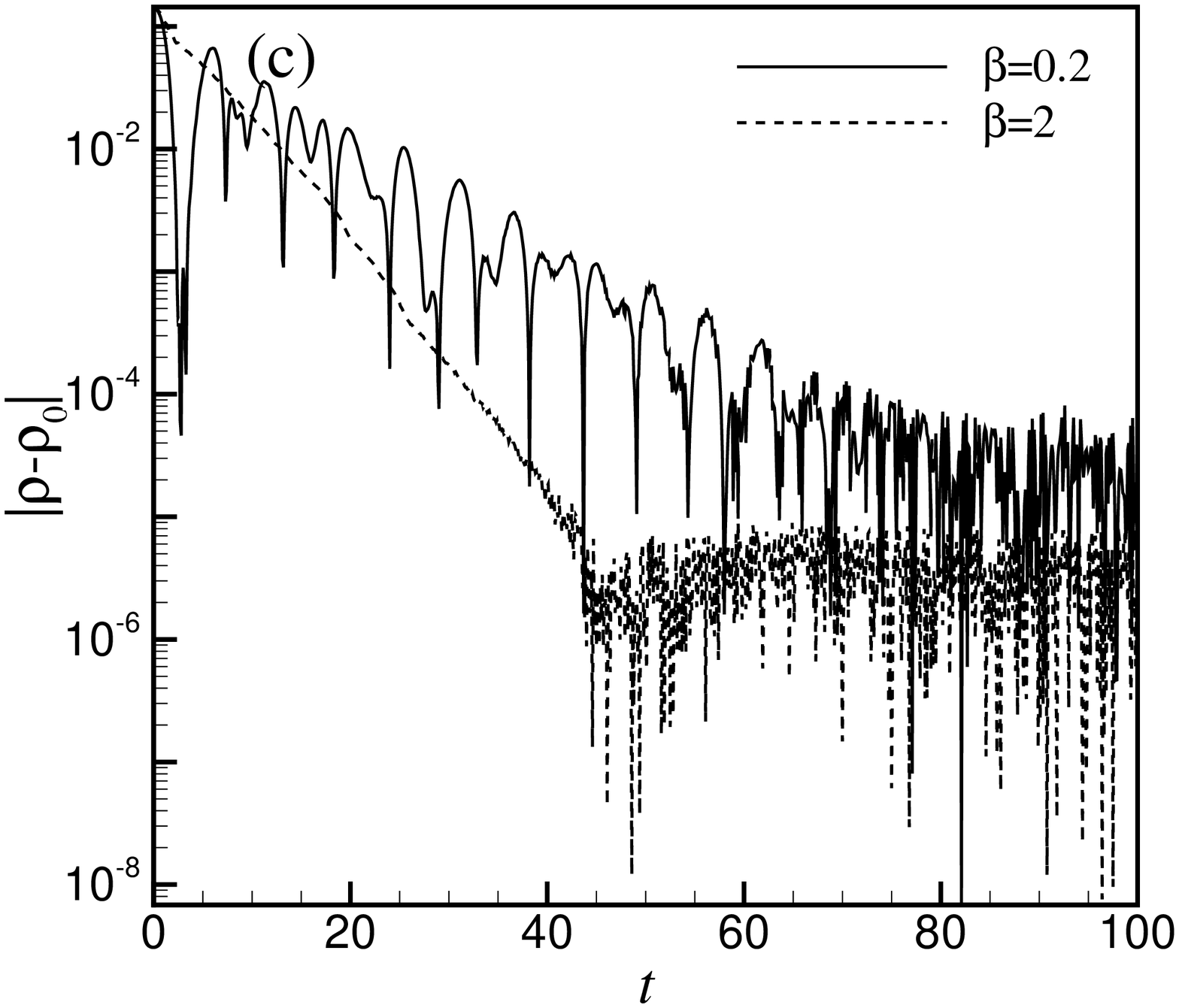} }  }

\caption{Evolution of $\rho_1(x,t)$ obtained through solving the Fokker-Planck equation. 
The initial DF is computed from equation (\ref{eq:initial-DF-stability}). Panels (a) and (b) correspond to 
$\beta=0.2$ and $\beta=2$, respectively. $\Delta x$ shows the travel range of the flock as the 
carrier fluid velocity varies according to $u_{\rm f}(t)=0.3 \cos(\omega t)$. Panel (c) illustrates 
the variation of $\vert \rho_1(x,t) \vert$ at $\xi=L/2$.}
\label{fig4_2}
\end{figure}

To this end, we show that a linearly stable mode of the perturbed continuity and momentum 
equations is also a stable solution of the full nonlinear Fokker-Planck equation. We set 
$a_1=-a_2=0.5 \Omega_1\Omega_2/(\Omega_2-\Omega_1)$ and start from the initial DF
\begin{eqnarray}
f(x,v,0)=\left [ \tilde \rho_0 + a_1\left ( \tilde \rho_{11} - \tilde \rho_{12}  \right ) \right ] \delta(v),
\end{eqnarray}
to numerically solve equation (\ref{eq:kinetic}) and trace the transient dynamics of the 
second discrete mode (with $\lambda_2=0.90$) in the model of Figure \ref{fig4}. 
In the linear regime, this mode decays monotonically for $\beta \ge 1.89$ and becomes an 
underdamped oscillatory wave otherwise. 
Figure \ref{fig4_2} shows several snapshots of $\rho_1(x,t)=\rho(x,t) - \rho_0(x,t)$ for $\beta=0.2$ 
and $\beta=2.0$ where $\rho_0(x,t)=\tilde \rho_0(\xi(x,t))$ is the traveling base-state density of particles. 
We have set the velocity of the carrier fluid to $u_{\rm f}(t)=0.3 \cos (\omega t)$ with $\omega = 0.94$, 
which is equal to $\rm {Im} (\Omega_{1,2})$ when $\lambda_2=0.90$ and $\beta=0.2$. 
Figure \ref{fig4_2}(a) demonstrates that the traveling perturbation is damped while it oscillates. 
No oscillation is seen in Figure \ref{fig4_2}(b) for $\beta=2$ as is expected. Although the flock 
velocity is asynchronous with $u_{\rm f}(t)$, the particle density has ultimately acquired its invariant 
traveling form $\tilde\rho_0(\xi)$. This is in agreement with our analytical results. In Figure \ref{fig4_2}, 
$\Delta x$ indicates the travel range of the entire flock as it responds to the pulsed flow. Our results 
show that after four oscillation cycles of $u_{\rm f}(t)$, $\vert \rho_1(x,t) \vert$ drops to $3.4\%$ 
and $0.1 \%$ of its initial value for $\beta=0.2$ and $\beta=2$, respectively. Figure \ref{fig4_2}(c) 
shows the temporal variation of $\vert \rho_1(x,t) \vert$ at $\xi=L/2$. The oscillatory behavior 
for $\beta=0.2$, and overdamped response for $\beta=2$ are evident in Figure \ref{fig4_2}(c). 
In both cases the envelope (or wave amplitude) decays linearly in the logarithmic scale, except 
in regions dominated by numerical noise as $\vert \rho_1(x,t)\vert \rightarrow 0$.

\section{Conclusions}
\label{sec:conclusions}

We studied the collective dynamics of interacting particles in steady and unsteady 
one-dimensional flows. One dimensional swarms of particles can be observed in 
microchannels when the sizes of particles and channel cross section are comparable. 
For instance, the collective motion of RBCs in micro vessels is a one-dimensional 
problem, especially when RBCs take a parachute shape \cite{MNG09}. Track cycling 
is another prominent one-dimensional problem where the members of a team 
collaborate along their path to minimize the drag force. Particle--particle 
interaction was modeled using the Morse potential. It was assumed that the carrier 
fluid is not disturbed by the particle phase although particles communicate with the 
carrier fluid through drag force. The evolution of the phase space distribution function 
of particles towards $H$-stable and catastrophic phases was investigated by keeping
inertial effects. These phases had been studied previously \cite{Lev09,ber10,Lev01}, 
but in the absence of acceleration terms in momentum equations.

One of the fundamental results of this study is the existence of two general families of 
discrete and singular decaying modes, which are calculated from a Fredholm-type 
eigenvalue problem. Singular modes are local disturbances supported by interactions 
of immediate neighbors, i.e., the information of local changes in the particle density is 
not communicated with the entire flock. Singular modes found here can be represented 
as the weighted sums of Dirac's delat functions, and they are analogous to van Kampen 
modes \cite{kampen55} in plasma oscillations. They are expected to form a complete 
set and be superposed to reconstruct a wide variety of density waves. Discrete modes, 
however, globally influence the flock and can be regarded as secondary collective 
effects developed in the vicinity of $\tilde\rho_0(\xi)$. It is noted that singular modes 
are the characteristics of systems of interacting particles and they {\it do not exist} 
in fluids.

Our findings have interesting implications for the flock of birds and fishes whose 
skins and body shapes have evolved to minimize drag force: in low-drag limits, 
discrete modes and superpositions of singular modes appear as long-lived, 
oscillatory density waves that travel through the entire population without 
disintegrating the flock. The only prerequisite for such a behavior is initially 
being in the catastrophic phase. Therefore, the collective flock of birds and fishes 
can exhibit a rich set of stable time-varying patterns, which should be observable 
during their group motions. 

Whether discrete and singular modes exist in two- and three-dimensional systems 
is an open problem. The difficulty of higher dimensional systems is to analyze the 
deformations of contact lines/surfaces between the particle and fluid phases 
(this problem does not exist in one-dimensional flows). In two dimensional systems 
a rotating vortex can emerge as the steady state solution \cite{Ors06,Lev01}. An interesting 
unsolved problem is to understand the linear stability of vortices. For doing so, 
one can adopt the usual polar coordinates $(R,\phi)$ to describe the 
governing equations. It is then possible to represent azimuthal perturbations 
using the Fourier series of $\phi$, and formulate the eigenvalue problem in 
terms of unknown functions that describe the radial variations of the particle 
density and two streaming velocity components. Singular two-dimensional modes 
may be produced from exact resonances between the radial and azimuthal motions 
of particles. Discrete modes, regardless of their stable or unstable nature, are also 
possible in the form of traveling spiral waves, analogous to density waves in 
self-gravitating stellar systems whose dynamics is represented by 
the collisionless Boltzmann equation \cite{J07}.

\section*{Acknowledgments}
We thank anonymous referees for their useful comments, which helped us to 
substantially improve the presentation of the paper.

\end{document}